\documentclass[aps, pra, reprint, superscriptaddress,longbibliography]{revtex4-2}

\usepackage{amsmath}   
\usepackage{graphicx}  
\usepackage{hyperref}  
\usepackage{lineno}    

\usepackage{lineno,hyperref}
\usepackage{eurosym}
\usepackage{amsmath,amssymb,graphicx}
\usepackage{color}
\usepackage[caption = false]{subfig}
\usepackage{booktabs}
\usepackage{multirow}
\usepackage{siunitx}

\usepackage{tikz}
\usetikzlibrary{arrows.meta, positioning}

\modulolinenumbers[5]
\newcommand{\bitem}{\begin{itemize}}
\newcommand{\fitem}{\end{itemize}}
\newcommand{\beq}{\begin{equation}}
\newcommand{\eeq}{\end{equation}}
\newcommand{\beqa}{\begin{eqnarray}}
\newcommand{\eeqa}{\end{eqnarray}}

\newcommand{\ket}[1]{|#1\rangle}                                

\newcommand{\UFSCar}{Departamento de Física, Universidade Federal de São Carlos, Rodovia Washington Luís, km 235 - SP-310, 13565-905 São Carlos, SP, Brazil}

\usepackage{orcidlink}

\begin{document}

\title{Quantum Features of the Thermal Two-Qubit Quantum Rabi Model in Ultra- and Deep-Strong Regimes}

\author{Ciro Micheletti~Diniz~\orcidlink{0000-0002-7602-0468}}\email{ciromd@outlook.com.br}
\affiliation{\UFSCar}

\author{Gabriella G. Damas~\orcidlink{0000-0003-3376-9281}}
\affiliation{Instituto de Física, Universidade Federal de Goiás, 74.001-970, Goiânia
- GO, Brazil}

\author{Norton G. de Almeida~\orcidlink{0000-0001-8517-6774}}
\affiliation{Instituto de Física, Universidade Federal de Goiás, 74.001-970, Goiânia
- GO, Brazil}

\author{Celso J. Villas-Boas~\orcidlink{0000-0001-5622-786X}}
\affiliation{\UFSCar}

\author{G. D. de Moraes Neto~\orcidlink{0000-0003-4273-8380}}
 \email{gdmneto@gmail.com}
\affiliation{College of Physics and Engineering, Qufu Normal University, Qufu, 273165, China}


\begin{abstract}
Quantum correlations and non-classical states are indispensable resources for advancing quantum technologies, and their resilience at finite temperatures is crucial for practical experimental implementations. The two-qubit quantum Rabi model (2QQRM), a natural extension of the quantum Rabi model, describes two qubits coupled to a single bosonic mode and has been extensively studied in cavity quantum electrodynamics, superconducting circuits, and quantum information science. In this work, we investigate the persistence of quantum correlations and non-classical states in the 2QQRM at thermal equilibrium, focusing on the ultrastrong and deep strong coupling regimes. Through a systematic analysis of quantumness quantifiers, we demonstrate the emergence of long-lived quantum correlations, even in the presence of thermal noise. Notably, we uncover striking phenomena arising from the interplay between detuning and deep strong-coupling: in the high-frequency limit, where the qubit energy exceeds the cavity-mode energy, quantum criticality emerges, leading to a high degree of photon squeezing. In contrast, the opposite regime is characterized by robust qubit-qubit quantum correlations. Importantly, we show that both dispersive regimes exhibit quantum features that are remarkably robust to parameter fluctuations, making them advantageous for maintaining quantum coherence. These results highlight the exceptional resilience of quantum resources in the 2QQRM and provide valuable insights for developing quantum technologies operating under realistic, finite-temperature conditions.
\end{abstract}

\maketitle

\section{Introduction}

The concept of quantumness, or nonclassicality, refers to the unique traits of quantum systems that cannot be fully explained by classical theories, whether deterministic or probabilistic. These traits include phenomena such as entanglement, coherence, and superposition, all of which challenge classical interpretations of reality. Quantum correlations in nonclassical states have become a foundational aspect of modern quantum information theory~\cite{chitambar2019quantum,streltsov2017colloquium}, with far-reaching implications for both theoretical research and the development of practical quantum technologies~\cite{acin2018quantum,nielsen2010quantum}. These correlations enable quantum information tasks such as secure communication~\cite{gisin2007quantum,QKD2009Review}, quantum sensing~\cite{degen2017quantum,Giovannetti2011,Giovannetti2006,Giovannetti2004,montenegro2023quantum,montenegro2022probing,montenegro2020mechanical}, and quantum simulation~\cite{Simulation2014Review}. At the same time, understanding the dynamics of quantum systems is crucial for the advancement of quantum technologies, where the effects of decoherence~\cite{zurek2003decoherence} and the transition from quantum to classical~\cite{modi2012classical} remain significant challenges. Techniques such as decoherence-free subspaces~\cite{mundarain2007decoherence,lidar1998decoherence,lidar2000protecting}, dynamical decoupling~\cite{viola2005random,celeri2008switching}, and reservoir engineering~\cite{reiter2016scalable,de2017steady, prado2009, de2014steady} have been proposed to mitigate these challenges and extend the lifetimes of nonclassical states.

Among the various platforms for generating and manipulating nonclassical states, quantum systems with light-matter interactions stand out as key candidates~\cite{blais2020quantum, wendin2017quantum}. Recent advances in the ultrastrong-coupling (USC)~\cite{anappara2009signatures, niemczyk2010circuit, forn2019ultrastrong} and deep strong-coupling (DSC)~\cite{yoshihara2017superconducting, yoshihara2018inversion} regimes have enabled the exploration of nonclassical behaviors even in the ground state of light-matter systems. In contrast to the strong-coupling (SC) regime~\cite{meschede1985one, thompson1992observation, weisbuch1992observation, lodahl2015interfacing, gu2017microwave}, where the coupling strength is comparable to or larger than the system decay rates but extremely small compared to system frequencies, the USC and DSC regimes exhibit a rich array of phenomena, including squeezing, entanglement, and photon antibunching~\cite{ashhab2010qubit, shen2014ground, ridolfo2013nonclassical, garziano2017cavity}. These effects arise due to the increased coupling strength, which becomes comparable to the system frequencies, allowing for more pronounced quantum correlations.

The quantum Rabi model (QRM), describing the interaction between a two-level atom (qubit) and a single-mode cavity field, serves as the fundamental framework for studying light-matter interactions~\cite{rabi1936process,rabi1937space,braak2011integrability,braak2016semi}. This model has been widely studied across various fields, such as quantum optics~\cite{scully1997quantum}, entanglement generation~\cite{chen2010entanglement}, quantum thermodynamics~\cite{altintas2015rabi,barrios2017role,alvarado2018quantum,wang2024critical,xu2024exploring}, and the study of quantum phase transitions (QPTs)~\cite{ashhab2013superradiance,hwang2015quantum,liu2017universal,hwang2016recurrent,puebla2016excited,hu2023excited,Liu2024using}. Modifications to the QRM, such as the anisotropic quantum Rabi model (AQRM)~\cite{xie2014anisotropic,tomka2014exceptional} and the Jaynes-Cummings model (JCM)~\cite{jaynes1963comparison}, provide even richer dynamics by incorporating different coupling terms and interactions. Both the AQRM and the JCM are known to exhibit first-order quantum phase transitions, which influence quantum thermalization and the persistence of quantum correlations at thermal equilibrium~\cite{fan2020quantum,liu2023quantum,ye2023implication,xu2024persisting}.

This work builds upon the two-qubit quantum Rabi model (2QQRM)~\cite{CS,PJB,wang2014solutions,LJM} and explores persisting quantum effects in light-matter systems operating in the USC and DSC regimes at thermal equilibrium. This model has been investigated in cavity quantum electrodynamics (QED)~\cite{SH}, superconducting circuit QED~\cite{LI,GT,RGL}, quantum information science~\cite{JA}, and cold atom physics~\cite{EL}. Applications range from preparing entangled atomic states in cavity QED~\cite{FJ,DL} to realizing two-qubit logic gates and enabling coherent storage and transmission in quantum information systems~\cite{SW,MBJ,SM}.
The 2QQRM has been analyzed using several methods, including exact approaches such as Bargmann-space techniques~\cite{PJA,PJB} and various approximations like the adiabatic method~\cite{LJM}, perturbation theory~\cite{CS}, the extended coherent state method~\cite{CQH}, the zeroth-order approximation~\cite{LJM}, and the generalized rotating-wave approximation (GRWA)~\cite{ZYY}. 

Regarding the previous techniques, the zeroth-order approximation is applicable when the qubit transition frequency $\Delta$ is much smaller than the field frequency $\omega$, even in the ultrastrong coupling regime. Additionally, in contrast to the rotating-wave approximation (RWA) that neglects counter-rotating terms and is valid to small couplings $g \ll \omega$ and detunings following $\left| \omega - \Delta \right| \ll \left| \omega + \Delta \right|$, the GRWA extends to negative detunings $\left( \delta = \omega - \Delta < 0 \right)$ and ultrastrong coupling $\left( g <\omega \right)$~\cite{ZYY}. These techniques have been employed to determine the energy spectrum, dynamics, and entanglement evolution of the system. Notably, for the 2QQRM, there exists a special dark state with at most one photon and constant eigenenergy~\cite{rodriguez2014searching}, where two identical qubits form a spin singlet, and the eigenstates are just products of these singlets and a Fock state. This dark state enables the deterministic fast generation of W states~\cite{peng2021one} and high-quality single-photon sources~\cite{peng2023deterministic} via adiabatic evolution. Recently, thermally-induced coherences in the 2QQRM have been investigated~\cite{abari2024correlated}. Additionally, new extensions of the model have been explored, including a two-qubit Rabi model incorporating explicit spin-spin interactions~\cite{grimaudo2024thermodynamic} and a tripartite quantum Rabi model, where a single bosonic mode couples to two (1/2)-spin particles via a spin-spin interaction, resulting in spin-spin-boson coupling~\cite{hamlyn2024tripartite}.

A central question arises in the context of the 2QQRM: can long-lived, highly correlated nonclassical states persist even in the presence of a thermal reservoir? This issue is crucial for the development of current experimental platforms where maintaining quantum coherence at finite temperatures is a significant challenge.

In this work, we address the aforementioned question by analyzing the \textit{quantumness}  of the global system steady state and deepen our studies in the finite temperature-coupling strength phase diagram considering broader ranges of parameters. To this end, the paper is structured as follows: In Sec.~\ref{The model}, we present the model and revisit some  
well-known results for the closed system scenario. In Sec.~\ref{Quantum Correlation and Nonclassicality measures}, we define the nonclassicality measures and witnesses used throughout our analysis. In Sec.~\ref{Results and discussion}, we present our findings, focusing on the temperature-dependent behavior of the system. Finally, Sec.~\ref{Conclusions} contains our concluding remarks.

\section{The Model}\label{The model}

\begin{figure*}
\begin{center}
\includegraphics[width=1\textwidth]{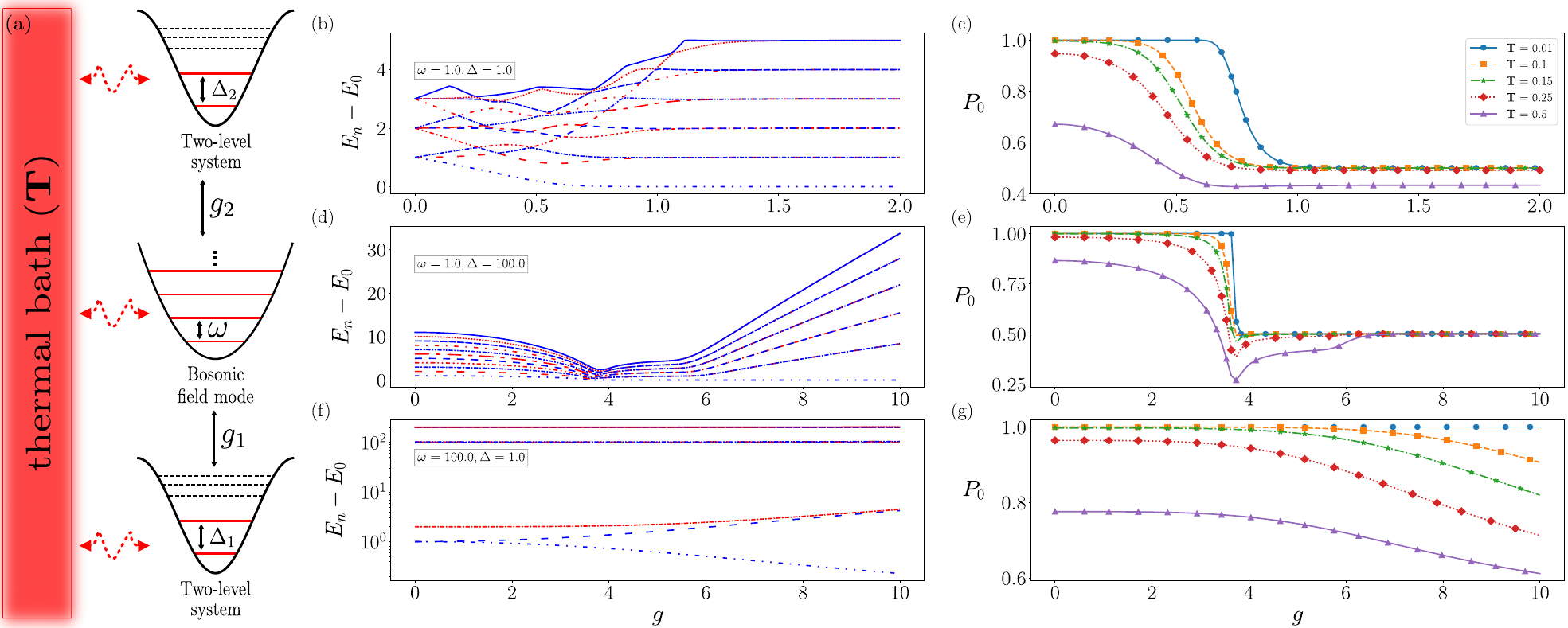}
\caption{Nonclassical effects in thermal Two-Qubit Quantum Rabi Model. Panel (a) illustrates the model studied in this work. A bosonic field mode of frequency $\omega$ interacts with two atoms, modeled as two-level systems. The $i$-th atom couples to the field mode with strength $g_i$ and has a transition frequency $\Delta_i$. All components are in thermal equilibrium with a shared bath at temperature \textbf{T}. Additionally, we present the energy difference $E_n - E_0$ between the $n$-th excited state (in increasing order, starting with $n = 1$) and the ground state considering the total system, along with the equilibrium Gibbs ground-state population $P_0$, as functions of the qubit-field coupling strength $g$ for degenerate qubits. We present three different interaction regimes: panels (b) and (c) represent the resonant case with $\omega = \Delta = 1.0$, panels (d) and (e) depict the high-frequency qubit regime where $\Delta = 100.0$, $\omega = 1.0$, and, finally, panels (f) and (g) correspond to the adiabatic quantum oscillator regime with $\omega = 100.0$, $\Delta = 1.0$. In the energy plots, states with different parities are distinguished using blue and red lines. Still, the population distributions are computed for a range of temperatures, with all quantities expressed in units of $\omega$ in panels (b)–(e) and in units of $\Delta$ in panels (f) and (g).}\label{EnPn}
\end{center}
\end{figure*}

The Hamiltonian of the two-qubit quantum Rabi model (2QQRM), as illustrated in Fig.~\ref{EnPn}(a), is given by ($\hbar = 1$)~\cite{CS,PJB}:
\begin{equation} \label{eq:Ham}
\hat{H} = \omega \hat{a}^{\dagger} \hat{a} + \sum_{i=1}^2 \frac{\Delta_i}{2} \hat{\sigma}_z^{(i)} + \sum_{i=1}^2 g_i (\hat{a} + \hat{a}^{\dagger}) \left( \hat{\sigma}^{(i)+} + \hat{\sigma}^{(i)-} \right),
\end{equation}
\noindent where the quantum system consists of two qubits (represented by the Pauli matrices $\hat{\sigma}_x, \hat{\sigma}_y, \hat{\sigma}_z, \hat{\sigma}^{\pm}$) interacting with a single-mode bosonic field. In this model, $\hat{a}$ and $\hat{a}^\dagger$ are the boson annihilation and creation operators with frequency $\omega$. The coupling strength $g_i$ describes the interaction between the $i$-th qubit and the bosonic field, while the transition frequencies of the two qubits are, respectively, $\Delta_1$ and $\Delta_2$.

One distinctive feature of the 2QQRM is its parity symmetry $\mathbb{Z}_2$, akin to the standard quantum Rabi model~\cite{forn2019ultrastrong}. This symmetry can be identified by examining the total excitation number operator $\hat{n} = \hat{a}^\dagger \hat{a} + \hat{\sigma}^{(1)+} \hat{\sigma}^{(1)-} + \hat{\sigma}^{(2)+} \hat{\sigma}^{(2)-}$ and the parity operator $\hat{\pi} = \exp(i\pi \hat{n})$. The 2QQRM satisfies the commutation relations $[\hat{H}, \hat{n}] \neq 0$ and $[\hat{H}, \hat{\pi}] = 0$. As a result, two important consequences follow: (i) the expected number of excitations in both the qubits and the bosonic mode is non-zero, even in the ground state, and (ii) $\hat{\pi}$ has eigenvalues $\langle \hat{\pi} \rangle = \pm 1$, depending on whether the total excitation number is even or odd. The Hilbert space can be naturally divided into triplet and singlet sectors. The triplet sector consists of the states $\left\{ \vert \Phi_{\pm, n} \rangle, \vert \Psi_{+,n} \rangle \right\}$, while the singlet sector contains the states $\left\{ \vert \Psi_{-,n} \rangle \right\}$. The states in these sectors are defined as:
\begin{align}
        \vert \Phi_{\pm, n} \rangle & = \frac{1}{\sqrt{2}} \vert n \rangle \left( \vert g,g \rangle \pm \vert e,e \rangle \right); \\
        \vert \Psi_{\pm,n} \rangle & = \frac{1}{\sqrt{2}} \vert n \rangle \left( \vert g,e \rangle \pm \vert e,g \rangle \right),
\end{align}
\noindent where $\ket{n}$ is the Fock state of $n$ photons, and $\ket{g}$ ($\ket{e}$) is the ground (excited) state of the qubits.


In the singlet sector, the eigenstates with an odd photon number $n$ belong to the positive parity subspace, while those with an even photon number $n$ belong to the negative parity subspace. This symmetry allows the analytical solution of the 2QQRM using methods such as the Bargmann-Fock space~\cite{PJA,PJB} of analytic functions and the Bogoliubov operator approach~\cite{chen2012exact}, which are also used for the standard QRM.

To study the quantum correlations and quantum nonclassicality in the 2QQRM, we aim to examine the system at thermal equilibrium. In this context, the system interacts with a reservoir at temperature \textbf{T}, and we seek to obtain the steady state of the system under these conditions. To ensure the steady state represents the true quantum state at thermal equilibrium for an arbitrary set of coupling strengths, one must adopt a Born-Markov master equation in a dressed picture~\cite{BRE02,xu2024persisting}. 
In the dressed-state picture, quantum jumps occur between the eigenstates of the full Hamiltonian, which accounts for both the system and its interaction with the reservoir. This contrasts with the standard master equation approach, where jumps are determined solely by the system’s free energy.

By considering conditions that lead to thermal equilibrium—such as when the qubits and the mode share the same temperature, $T_{\text{mode}} = T_{\text{qubits}} = \textbf{T}$, or when only a single subsystem is coupled to the reservoir—the system's dynamics can be effectively described using the dressed master equation. The dissipative evolution is governed by
\begin{eqnarray}
	\frac{d}{dt}\hat{\rho}&=&-i[\hat{H},\hat{\rho}]+\sum_{ \substack{u=a,\sigma^{(i)}\\k<j } }
	\{\Gamma^{jk}_un_u(\Delta_{jk})\mathfrak{D}[|\phi_j{\rangle}{\langle}\phi_k|,\hat{\rho}]\nonumber\\
	&&+\Gamma^{jk}_u[1+n_u(\Delta_{jk})]\mathfrak{D}[|\phi_k{\rangle}{\langle}\phi_j|,\hat{\rho}]\}=\mathcal{L}(t)\hat{\rho}, \label{eq:dressed-me}
\end{eqnarray}
\noindent where $n_u(\Delta_{jk})$ is the number of thermal photons of the $u$-th bath associated with the energy gap $\Delta_{jk}=E_{j}-E_{k}$, $\ket{\phi_k}$ and $E_k$ are, respectively, the eigenstates and eigenenergies of the full Hamiltonian $\hat{H}$, satisfying $\hat{H}\ket{\phi_k} = E_k\ket{\phi_k}$, and $\mathcal{L}(t)$ is the Liouvillian superoperator, with the Lindblad dissipator given by
\begin{equation} \mathfrak{D}[\hat{O},\hat{\rho}]=\frac{1}{2}[2\hat{O}\hat{\rho}\hat{O}^{\dag}-\hat{\rho}\hat{O}^{\dag}\hat{O}-\hat{O}^{\dag}\hat{O}\hat{\rho}]. \end{equation}
\noindent 
In Eq.~\eqref{eq:dressed-me}, we also define the dissipation rates 
\begin{equation}
\Gamma^{jk}_u=\gamma_u(\Delta_{jk})|S^{jk}_u|^2,
\end{equation}
which consist of two distinct contributions. The first is the spectral function,
\begin{equation}
\gamma_u(\Delta_{jk})=2\pi\sum_k|\lambda_{k,u}|^2\delta(\Delta_{jk}- \omega_k),
\end{equation}
where $\lambda_{k,u}$ accounts for the $u$-th thermal bath coupled to a single-mode boson field with the frequency $\omega_k$. The second contribution comes from the transition coefficients of the $i$-th qubit ($q$) and the cavity mode ($c$), respectively, given by
\begin{eqnarray}
    {S}^{jk}_{q_{(i)}} &=& {\langle}\phi_j|(\hat{\sigma}^{(i)+}+\hat{\sigma}^{(i)-}|\phi_k{\rangle},\\
    {S}^{jk}_c &=& {\langle}\phi_j|(\hat{a}^\dag+\hat{a})|\phi_k{\rangle}.
\end{eqnarray}
To satisfy the necessary conditions for the validity of the master equation presented in Eq.~\eqref{eq:dressed-me}, we consider the Ohmic case:
\begin{equation}
    \gamma_u(\Delta_{jk})=\pi\alpha\Delta_{jk} e^{-|\Delta_{jk}|/\omega_{c}},
\end{equation}
where $\alpha$ is the coupling strength between the system and the environment and $\omega_{c}$ is the cutoff frequency of thermal baths. Throughout all numerical simulations performed, we consider $\alpha=0.001\omega$ and $\omega_{c}=10\omega$. We note that for Gibbs states of the reservoir at temperature \textbf{T}, the correlation functions also obey the Kubo-Martin-Schwinger (KMS) relations~\cite{BRE02}, i.e.,  $\gamma_u(-\Delta_{jk})=\gamma_u(\Delta_{jk})e^{-\Delta_{jk}/k_B \textbf{T}}$, where $k_B$ is the Boltzmann constant.
Under these equilibrium conditions, the system reaches a steady state described by the canonical ensemble, with a density matrix ($k_B = 1$)
\begin{eqnarray}
	\hat{\rho}_\mathrm{ss}=\sum_k\frac{e^{-E_k/\textbf{T}}}{\mathcal{Z}}|\phi_k{\rangle}{\langle}\phi_k|,\label{eq:ss}
\end{eqnarray}
\noindent where $\mathcal{Z} = \sum_k e^{-E_k/\textbf{T}}$ is the partition function. The steady-state populations are then given by
\begin{equation} P_k = \frac{e^{-E_k/\textbf{T}}}{\mathcal{Z}}, \end{equation}
\noindent where the partition function $\mathcal{Z}$ guarantees that the total population sums to one, ensuring the proper statistical normalization.

To enhance the clarity of the numerical results, we follow by providing a concise review of significant limit regimes of our model.

Starting by the dispersive limit, where the qubits are far detuned from the resonator compared to the coupling strength, we have the condition $g\sqrt{n+1} \ll |\Delta - \omega| \ll |\Delta + \omega| $, where $n = \langle \hat{a}^\dagger \hat{a} \rangle$ is the number of photons in the resonator. The first inequality indicates the dispersive regime, where the coupling $g$ is small relative to the detuning, while the second condition is typically used to derive the rotating wave approximation Hamiltonian. Beyond the RWA, the detuning does not need to be smaller than both $\Delta$ and $\omega$, as long as $g \ll |\Delta - \omega|=\delta$ holds.

For two qubits, considering the RWA, the dispersive Hamiltonian is~\cite{zueco2009qubit} 
\begin{equation}
\begin{aligned} 
\hat{H}_{\text{disp}} & = \omega\hat{a}^\dagger \hat{a} + \frac{1}{2}\sum^{2}_{j=1}\left( \Delta_j + \frac{g^2_j}{\delta_j} \right)\hat{\sigma}^{(j)}_z \\ 
& + \sum^2_{j=1}  \frac{g^2_j}{\delta_j} \hat{a}^\dagger \hat{a}\hat{\sigma}^{(j)}_z + J_{12}\left( \hat{\sigma}^{(1)}_+ \hat{\sigma}^{(2)}_- + \hat{\sigma}^{(1)}_- \hat{\sigma}^{(2)}_+  \right),
\end{aligned}\label{H disp}
\end{equation}
\noindent where $J_{12}$ is the effective coupling between the two qubits, given by
\begin{equation}
J_{12} = g_1 g_2 \left( \frac{1}{\delta_1} + \frac{1}{\delta_2} \right),
\end{equation}
\noindent with $\delta_i = \Delta - \omega_i$ being the detuning between the $i$-th qubit and the field mode.

When extending this Hamiltonian beyond the RWA, we obtain
\begin{equation}
\begin{aligned}
\hat{H}_{\text{disp}} & =  \hbar \omega\hat{a}^\dagger \hat{a} + \frac{1}{2}\sum^{2}_{j=1}\Delta_j \hat{\sigma}^{(j)}_z + \overline{J}_{12} \hat{\sigma}^{(1)}_x \hat{\sigma}^{(2)}_x \\ 
& + \sum^2_{j=1} g^2_j \left( \frac{1}{\delta_j} - \frac{1}{\Delta_j - \delta_j} \right) \left(\hat{a}^\dagger  + \hat{a} \right)^{2}\hat{\sigma}^{(j)}_z,
\end{aligned}\label{eq:Heff}  
\end{equation}
\noindent where the modified coupling strength is
\begin{equation}
\overline{J}_{12} = g_1 g_2 \left( \frac{1}{\delta_1} + \frac{1}{\delta_2} - \frac{1}{2\Delta_1-\delta_1} - \frac{1}{2\Delta_2-\delta_2} \right).
\end{equation}

In the case of two qubits, the non-RWA treatment yields a qualitatively different effective model compared to the RWA. Specifically, the interaction between the qubits becomes of the Ising type $\left( \hat{\sigma}^{(1)}_x \hat{\sigma}^{(2)}_x \right)$ in the non-RWA regime, while in the RWA, it corresponds to an XY-type interaction $\left( \hat{\sigma}^{(1)}_+ \hat{\sigma}^{(2)}_- + \hat{\sigma}^{(1)}_- \hat{\sigma}^{(2)}_+  \right)$. In addition to this, the non-RWA Hamiltonian does not conserve the number of excitations, which has important implications for the design and implementation of two-qubit gates. More importantly, the coupling is no longer proportional to the photon number operator $\hat{a}^\dagger \hat{a}$, as in the RWA, but instead involves a parametric $\left(\hat{a}^\dagger  + \hat{a} \right)^{2}$ term. This modification opens the possibility for squeezing of the field, which can lead to nonclassical effects such as entanglement and photon-number squeezing, further complicating gate operations and providing richer dynamics for quantum information processing.
For two degenerate qubits ($\Delta_1 = \Delta_2$) coupled to a single cavity, the ground state of the non-RWA Hamiltonian is $|0\rangle \left( | g g\rangle - \frac{J}{2\Delta} |e e \rangle \right)$,
which exhibits qubit-qubit entanglement. In contrast, the corresponding ground state of the two-qubit RWA Hamiltonian is the product state $|0\rangle |g g\rangle$. These effective Hamiltonians shed light on the regime of coupling parameters where qubit-qubit quantum correlations arise.

In the deep strong coupling
regime, where the coupling parameters $g_1, g_2 \gg \Delta_1, \Delta_2$, the two-qubit quantum Rabi model
Eq.~\eqref{eq:Ham} can be approximated as a leading Hamiltonian~\cite{CS} $\hat{H} = \hat{H}_0 + \hat{P}$, where $\hat{H}_0$ is diagonal and $\hat{P}$ is the perturbation. When applying a general rotation $\hat{R}_y = e^{-i\frac{\pi}{4}\hat{\sigma}^{(1)}_y}\otimes e^{-i\frac{\pi}{4}\hat{\sigma}^{(2)}_y}$ 
that transform the qubit operators $\hat{\sigma}_x^{(j)}$ and $\hat{\sigma}_z^{(j)}$, the effective Hamiltonian becomes
\begin{align}
        \tilde{H}_0 & = \omega \hat{n} + \left( \hat{a} + \hat{a}^{\dagger} \right) \left( g_1 \hat{\sigma}_z^{(1)} + g_2 \hat{\sigma}_z^{(2)} \right), \\
        \tilde{P} & = -\frac{1}{2} \left( \Delta_1 \hat{\sigma}_x^{(1)} + \Delta_2 \hat{\sigma}_x^{(2)} \right),
\end{align}
\noindent where the rotated Hamiltonian becomes diagonal,
\begin{equation}
\tilde{H}_0 = \hat{T}_D \left[ \omega \hat{n} - \frac{1}{\omega} \begin{pmatrix} g_+^2 & 0 & 0 & 0 \\ 0 & g_-^2 & 0 & 0 \\ 0 & 0 & g_-^2 & 0 \\ 0 & 0 & 0 & g_+^2 \end{pmatrix} \right] \hat{T}_D^\dagger,
\end{equation}
for a driven oscillator basis $\hat{T}_D$ obeying
\begin{equation}
\hat{T}_D =  \begin{pmatrix} \hat{D}(g_+/\omega) & 0 & 0 & 0 \\ 0 & \hat{D}(g_-/\omega) & 0 & 0 \\ 0 & 0 & \hat{D}(-g_-/\omega) & 0 \\ 0 & 0 & 0 & \hat{D}(-g_+/\omega) \end{pmatrix},
\end{equation}
\noindent with the displacement operator $\hat{D}(\alpha) = e^{\alpha \hat{a}^\dagger - \alpha^* \hat{a}}$ and $g_{\pm} = g_1 \pm g_2$. 

Still, the approximated eigenvalues, including second-order corrections, exhibit a twofold degeneracy. The first-order correction due to the perturbation $\tilde{P}$ is null, while the second-order correction arises from photon transitions mediated by the displacement operator, depending explicitly on the coupling strengths $g_1$ and $g_2$. Once we are dealing with the DSC regime, where $g_1, g_2$ reach large values, these second-order corrections diminish, leading the spectral branches to converge toward the spectrum of a forced harmonic oscillator~\cite{CS}.

\begin{figure*}[t!]
\includegraphics[width=1.\linewidth]{diagrama2.pdf}
\caption{Regimes of the two-qubit quantum Rabi model. Key features: (1) RWA and non-RWA dispersive limits show fundamentally different entanglement structures, (2) deep strong coupling reveals displaced Fock product states, (3) the adiabatic regime maintains qubit correlations despite a harmonic spectrum, and (4) the high-frequency limit exhibits squeezing and superradiant-like behavior.}
\label{diagram}
\end{figure*}

The next limit we consider is where the oscillator's characteristic frequency $\omega$ is large compared to the qubit's energy splitting (i.e., $\omega \gg \Delta_i$) and also large compared to the coupling strength ($\omega \gg g_i$). In this case, the oscillator can be considered to remain in its initial energy eigenstate (e.g., the ground state, first excited state, etc.), and this state evolves adiabatically in response to changes in the qubit's state.

It has been shown in Ref.~\cite{agarwal2012tavis} that in the regime $\Delta_i \leq 0.25\omega$, the adiabatic approximation holds, and the eigenstates of the system are product states of displaced Fock states for the photons and atomic states that are $x$-polarized, i.e., $\ket{\Phi} = \ket{j,m}\ket{\phi_{m}}$. Specifically, the qubits can be described using the simultaneous eigenstates of the operators $\hat{\sigma}_x^{(1)}$ and $\hat{\sigma}_x^{(2)}$, where $\hat{\sigma}_x^{(i)} \ket{\pm} = \pm 1 \ket{\pm}$. In terms of these eigenstates, the states $\ket{j,m}$ of the collective spin operator $\hat{S}_x = \frac{1}{2}(\hat{\sigma}_{x}^{(1)} + \hat{\sigma}_{x}^{(2)})$ can be expressed as:
\begin{equation}\begin{pmatrix}\label{e.unit_trans} \ket{1,1} \\ \ket{1,0} \\ \ket{0,0} \\ \ket{1,-1} \end{pmatrix}
=
\begin{pmatrix} 
1 & 0 & 0 & 0 \\ 
0 & \frac{1}{\sqrt{2}} & \frac{1}{\sqrt{2}} & 0 \\
0 & \frac{1}{\sqrt{2}} & -\frac{1}{\sqrt{2}} & 0 \\
0 & 0 &  0 & 1 
\end{pmatrix}
\begin{pmatrix} \ket{+,+} \\ \ket{+,-} \\ \ket{-,+} \\ \ket{-,-} \end{pmatrix},\end{equation}
\noindent where $j \in \{0,1\}$ and $m \in \{-1,0,1\}$. Additionally, $\ket{\phi_m}$ is the
Fock state displaced by $\hat{D}(-mg)$.

The adiabatic approximation highlights a significant degree of quantum correlations between the qubits in this regime, in agreement with the effective Hamiltonian in Eq.~\eqref{eq:Heff}, a conclusion that will be further substantiated by our forthcoming numerical results.

We now turn to the high-frequency limit, where the qubit adjusts adiabatically, i.e., $\Delta_i \gg \omega, g_i$. In this regime, quantum criticality emerges in QRMs and generalized QRMs~\cite{hwang2015quantum,liu2017universal,ashhab2013superradiance}, leading to abrupt changes at quantum phase transitions (QPT) due to quantum fluctuations. This phenomenon is exemplified by superradiance phase transitions (SPT), where excitations in the ground-state population occur~\cite{larson2017some}. Experimental ion trap studies~\cite{cai2021observation} have confirmed second-order QPT in the QRM, with a rapid increase in both phonon numbers and ion excitations at the critical point. In the deep strong coupling regime, a zero-temperature photon vacuum instability drives the system toward a QPT at the critical coupling strength $g_c = \frac{\sqrt{\omega \Delta}}{2}$~\cite{hwang2015quantum,liu2017universal,ashhab2013superradiance}. Below $g_c$, the system resides in a normal phase, while it enters a superradiant phase above $g_c$. In the Dicke model~\cite{wang1973phase,xu2024universal}, the thermodynamic limit is attained as the number of particles increases. However, in few-atom systems, quantum phase transitions emerge in the classical oscillator limit, characterized by $\omega/\Delta \rightarrow 0$. A distinct thermodynamic limit arises in the two-qubit quantum Rabi model with spin-spin interactions~\cite{grimaudo2024thermodynamic}, where criticality is determined by the infinite ratios of the spin-spin and spin-mode couplings to the mode frequency, independent of the spin-to-mode frequency ratios.
Although this work does not primarily focus on the system's critical behavior, our findings in the $\Delta \gg \omega$ regime suggest its presence, as evidenced by the photon population and their pronounced nonclassicality.

We also briefly investigated the Liouvillian spectrum $\mathcal{L}$~\cite{minganti2018spectral}. This analysis not only confirms the uniqueness of the equilibrium state described in Eq.~\eqref{eq:ss}, ruling out the possibility of a dissipative quantum phase transition, but also provides key insights into the system's relaxation timescale. Let $\{\mu_\alpha\}_{\alpha=0,1,2,\ldots}$ denote the complex eigenvalues of $\mathcal{L}$, as defined in Eq.~\eqref{eq:dressed-me}. In the parameter range analyzed in this work, numerical evaluation indicates that $\mu_0=0$ is non-degenerate, ensuring a unique and stable steady state. To quantify the relaxation timescale, we consider the Liouvillian gap~\cite{mori2023symmetrized}, defined as the difference between zero (i.e., $\mu_0=0$) and the eigenvalue with the largest nonzero real part:  $\mu_{1} = -\max_{\alpha\neq 0}\mathrm{Re}[\mu_\alpha]$. The Liouvillian gap determines the asymptotic decay rate and provides an estimate for the thermalization time, given by $t_{\text{th}} \approx |\mu_1|^{-1}$.
To compare the relaxation time required to reach equilibrium in the 2QQRM, we calculate the ratio of the normalized Liouvillian gap $\mu_1(\omega,\Delta,g)$ to the gap of the decoupled system, $\mu_1(\omega,\Delta,g=0)$. Our findings reveal:
(i) near resonance ($\omega/\Delta \approx 1$), the ratio falls within the range $1/2 < \mu_1(\omega,\Delta,g)/\mu_1(\omega,\Delta,g=0) < 2$;
(ii) in the limit $\omega \gg \Delta$, the gap remains nearly unchanged, $\mu_1(\omega,\Delta,g)/\mu_1(\omega,\Delta,g=0) \approx 1$;
(iii) for $\Delta \gg \omega$, the ratio can vary significantly, spanning $1 < \mu_1(\omega,\Delta,g)/\mu_1(\omega,\Delta,g=0) < 10^5$, with the upper bound corresponding to the extreme case of $\omega/\Delta = 10^{-3}$.
These results are consistent with the thermalization time in the thermal AQRM~\cite{xu2024persisting} and with Ref.~\cite{chen2024suppressed}, which reported that the energy relaxation rate decreases as the system approaches criticality and may even vanish precisely at the critical point. Notably, a faster thermalization is essential for quantum thermodynamic applications, as shorter relaxation times can enhance the power of quantum heat engines. Additionally, thermalization dynamics play a key role in enabling the on-demand generation of equilibrium non-classical states for quantum information processing.

 Summarizing the theoretical results for limit cases aforementioned, we elaborate the diagram presented in Fig.~\ref{diagram}, where we show in a concise way the differences and specifics of each parameter regime. In addition, numerical computations throughout this work were carried out using QuTiP~\cite{qutip2}. In all simulations, the full Hamiltonian of Eq.~\eqref{eq:Ham} was considered. To ensure accurate results without expending unnecessary computational labor, we use a variable Fock basis, checking its precision in each case by computing the bosonic commutator of the Fock annihilation and creation operators, which presents an error smaller than $10^{-4}$. Besides, by considering larger bases during preliminary tests, we verify the convergence of the results for all the quantities studied throughout this work.

To conclude, we present in Fig.~\ref{EnPn} the energy spectrum (with blue and red indicating different parity states) and steady-state population $P_0$ across different temperatures for degenerate qubits in various regimes. Figs.~\ref{EnPn}(b) and (c) show the resonant interaction regime ($\omega = \Delta$), where the energy levels exhibit significant anharmonicity, transitioning to a degenerate harmonic spectrum in the deep strong coupling limit. Figs.~\ref{EnPn}(d) and (e) illustrate the high-frequency qubit regime ($\Delta \gg \omega, g$), revealing distinct behaviors before and after the critical coupling $g_c = \frac{\sqrt{\omega \Delta}}{2}$. Figs.~\ref{EnPn}(f) and (g) show the adiabatic quantum oscillator regime ($\omega \gg \Delta, g$), where the system evolves adiabatically and the qubits track the oscillator's evolution in a nearly flat potential.

\section{Quantum Correlation and Nonclassicality Quantifiers}\label{Quantum Correlation and Nonclassicality measures}

To investigate how the system’s light-matter coupling influences thermal quantum correlations and thermal nonclassicality (i.e., quantum effects that persist even at thermal equilibrium) we focus on several key quantities that provide insights into these nonclassical properties. The quantifiers employed throughout this paper are designed to ensure a comprehensive description of the system, offering a multifaceted understanding of its quantum behavior. Specifically, we analyze:

\begin{enumerate}
    \item \textbf{The zero-delay second-order correlation function, $G^{(2)}(0)$}: This quantifies photon bunching or anti-bunching, highlighting deviations from classical photon statistics. It reveals quantum characteristics by indicating whether photons tend to arrive together (bunching) or separately (anti-bunching) at the detector, thus serving as a tool for identifying single-photon sources.

    \item \textbf{Bosonic field quadrature squeezing, $\zeta^{2}$}: This quantifies the squeezing of the electromagnetic field quadratures (such as position or momentum) below the vacuum state uncertainty limit whenever $\zeta^{2}<1$.

    \item \textbf{Negativity, $\mathcal{N}(\hat{\rho})$}: This quantifies the degree of separability in a quantum state by examining the eigenvalues of the partial transpose of the density matrix. Here, it is employed to study entanglement between photons and qubits in the system.

    \item \textbf{Mutual information, $\mathcal{I}(\hat{\rho})$}: This quantifies the total correlations, both classical and quantum, between subsystems in a quantum system, providing a complete measure of the shared information.

    \item \textbf{Concurrence, $C(\hat{\rho})$}:  This is an entanglement measure specific to two-qubit systems. It quantifies the extent to which a quantum state deviates from being separable. A concurrence value of zero indicates no entanglement between the two qubits.

    \item \textbf{Quantum discord, $\mathcal{D}(\hat{\rho})$}: This captures the difference between total correlations (as measured by mutual information) and classical correlations that can be extracted via local measurements. It provides a nuanced view of quantum correlations beyond entanglement.

    \item \textbf{Quantum coherence}, $C_\mathrm{RE}$: This reflects the ability of a quantum system to exhibit interference effects, measuring the presence of superposition states through the nonzero off-diagonal elements of the density matrix.

    \item \textbf{Local quantum uncertainty (LQU), $\mathcal{U}_A(\hat{\rho})$}: This quantifies the minimum uncertainty in local observables of a quantum system, offering insights into quantum features at the level of individual qubits.
\end{enumerate}

In the following subsections, we provide detailed definitions and interpretations of each of these quantities, highlighting their significance in studying thermal nonclassicality and the impact of light-matter coupling.

\subsection{Zero-delay second-order correlation function}

Sub-Poissonian photon statistics serve as a signature of the quantum nature of light, distinguishing it from classical light descriptions. While this property does not always manifest in all quantum states~\cite{carmichael1999statistical,carmichael2009statistical}, its presence signals nonclassicality. The second-order correlation function $g^{(2)}(\tau)$ classifies light as follows: (i) $g^{(2)}(\tau)>1$ for bunched light (classical chaotic), (ii) $g^{(2)}(\tau)=1$ for coherent light (classical random), and (iii) $g^{(2)}(\tau)<1$ for antibunched light (nonclassical)~\cite{carmichael1999statistical,carmichael2009statistical}.

In the USC regime, where strong light-matter interaction occurs, the zero-delay second-order correlation function is given by~\cite{rabl2011photon,ridolfo2012photon} \begin{equation}
    G^{(2)}(0) = \frac{\langle (\hat{X}^-)^2 (\hat{X}^+)^2 \rangle}{\langle \hat{X}^- \hat{X}^+ \rangle^2}, \label{eq:g2-x}
\end{equation}
\noindent where $ \hat{X}^+ = -i \sum_{k>j} \delta_{kj} X_{jk} |\phi_j\rangle \langle \phi_k| $ with $ \hat{X}^- = (\hat{X}^+)^\dagger $, $ \delta_{kj} = E_k - E_j $ being the energy gap, and $ X_{jk} = \langle \phi_j|(\hat{a}^\dagger + \hat{a})|\phi_k\rangle $. In the weak coupling limit, $ \hat{X}^+ $ reduces to $ \hat{X}^+ = -i \omega \hat{a} $, simplifying the correlation function to the standard form.

A key observation in the USC regime is that in the eigenbasis of the Hamiltonian, the dressed light-matter jump operator $\hat{X}^{+}$ yields the correct expression for the ground state's excitation number: $\langle \phi_0|\hat{X}^-\hat{X}^+|\phi_0 \rangle = 0$, in contrast to the seemingly incorrect result $\langle \phi_0|\hat{a}^\dagger \hat{a}|\phi_0 \rangle \neq 0$. The dressed ground state, expressed in the bare light-matter basis $\{|g\rangle, |e\rangle, |n\rangle \}$, contains virtual excitations~\cite{garziano2013switching}. To observe these as real photons, the coupling must be switched on and off, allowing the virtual excitations to be detected via the intracavity photon operator $\hat{a}$. This process, which should occur on the timescale of the system’s relevant frequencies $\omega$ and $\Delta$~\cite{garziano2013switching}, allows the correct photon count $\langle \phi_k|\hat{a}^\dagger \hat{a}|\phi_k \rangle \neq 0$ to be obtained. The photons in the ground state are virtual and cannot be detected unless the coupling is abruptly switched off~\cite{de2014steady}.

\subsection{Quadrature Squeezing}
The second-order correlation function, \( G^{(2)}(0) \), often fails to serve as a reliable indicator of nonclassicality for a set of quantum states of light. Specifically, nonclassical Gaussian states, such as squeezed states, can display photon bunching or antibunching depending on the modulation of amplitude fluctuations. Therefore, to supplement \( G^{(2)}(0) \) as a nonclassicality quantifier, we compute the degree of squeezing, which is more effective at revealing nonclassical behavior in these cases.

The degree of squeezing is computed by analyzing the generalized rotated field quadrature, defined as \( \sqrt{2} \hat{x}_{\theta} = \hat{a} e^{-i\theta} + \hat{a}^{\dagger} e^{i\theta} \), where different values of \( \theta \) correspond to distinct field quadratures. In particular, the quadratures \( \hat{x}_{\theta=0} \) and \( \hat{x}_{\theta=\pi/2} \) correspond to the position and momentum operators, respectively, which satisfy the commutation relations. The squeezing parameter is given by
\begin{equation}
\zeta^{2} = \underset{\theta \in (0, 2\pi)}{\text{min}} (\Delta \hat{x}_{\theta})^2,
\end{equation}
where the minimization is taken over all possible angles \( \theta \). A value of \( \zeta^{2} < 1 \) indicates squeezing, while \( \zeta^{2} = 1 \) corresponds to a coherent state.

Assuming the light-matter interaction has been switched off, the above expression simplifies to 
\begin{equation}
\zeta^{2} = 1 + 2 \langle \hat{a}^{\dagger} \hat{a} \rangle - 2 |\langle \hat{a}^2 \rangle|,
\end{equation}
where the expectation values are taken considering the steady state. 

It is worth noting that previous studies, such as Refs.~\cite{stassi2016output, noh2019output}, have explored the squeezing of output quadratures in the USC and DSC regimes using dressed light-matter jump operators for a vacuum input field. Interestingly, these studies found that for any open system in its ground state, output squeezing cannot be produced, even if the ground state itself is squeezed. Since most of the squeezing in the two-qubit Rabi model is present in the ground state, this generalized measure leads to a zero degree of squeezing in the thermal state when the light-matter interaction is active. For this reason, we focus on evaluating the degree of squeezing after turning off the light-matter interaction. This process effectively transforms virtual photons into real, measurable photons, revealing a clearer signature of squeezing in the system. Thus, after switching off the interaction,  we capture the squeezing behavior by computing \( \zeta^{2} \), highlighting the nonclassical features of the state. To effectively convert virtual excitations into real photons, the light-matter coupling must be switched non-adiabatically, on timescales shorter than the inverse of the system's characteristic frequencies~\cite{garziano2013switching}. Currently, switching times of a few nanoseconds -- corresponding to approximately $10$ – $100$ cycles of GHz frequencies -- are achievable~\cite{chen2014qubit}. While finite switching durations and decoherence could, in principle, introduce imperfections, the available experimental techniques are adequate for observing the predicted photonic effects.

\subsection{Negativity}
To investigate quantum correlations between the light-matter subsystems, we use the negativity $\mathcal{N}(\hat{\rho})$ as a measure of entanglement in bipartite systems~\cite{eisert1999comparison,zyczkowski1998volume,plenio2005logarithmic,lee2000partial,vidal2002computable}. The negativity is defined as
\begin{equation}
	\mathcal{N}(\hat{\rho})=\frac{\Vert\hat{\rho}^{T_{A}}\Vert_{1}-1}{2},
\end{equation}
where $\hat{\rho}^{T_{A}}$ is the partial transpose of the quantum state $\hat{\rho}$ with respect to subsystem A, and $\Vert \hat{Y}\Vert_{1}={\rm Tr}|\hat{Y}|={\rm Tr}\sqrt{\hat{Y}^{\dagger}\hat{Y}}$ denotes the trace norm, or the sum of the singular values of the operator $\hat{Y}$. Alternatively, the negativity can be computed as $\mathcal{N}(\hat{\rho})=\frac{1}{2}\sum_i(|\varepsilon_i|-\varepsilon_i)$, where $\varepsilon_i$ are the eigenvalues of the partially transposed density matrix $\hat{\rho}$. A negativity of $\mathcal{N}(\hat{\rho})=0$ indicates that the state is separable, meaning it is not entangled.

\subsection{Concurrence}
The concurrence is a commonly used measure of entanglement for bipartite quantum systems. For a given two-qubit state $\hat{\rho}$, the concurrence is defined as
\begin{equation}
C(\hat{\rho}) = \max\left(0, \lambda_1 - \lambda_2 - \lambda_3 - \lambda_4 \right),
\end{equation}
where $\lambda_i$ are the square roots of the eigenvalues of the matrix $\hat{\rho} \tilde{\hat{\rho}}$, with $\tilde{\hat{\rho}} = (\hat{\sigma}_y \otimes \hat{\sigma}_y) \hat{\rho}^* (\hat{\sigma}_y \otimes \hat{\sigma}_y)$, $\hat{\sigma}_y$ being the Pauli-Y matrix, and $\hat{\rho}^*$
is the matrix obtained by taking the complex conjugate of the elements of $\hat{\rho}$. The concurrence takes values between 0 and 1, with $C(\hat{\rho}) = 0$ indicating separable (non-entangled) states and $C(\hat{\rho}) = 1$ indicating maximally entangled states~\cite{wootters1998entanglement}.

\subsection{Mutual Information and Quantum Discord}

Quantum discord (QD) is a measure of quantum correlations that goes beyond entanglement. It is based on the difference between two classically equivalent expressions for mutual information when applied to quantum systems. This expressions for a bipartite system $\mathcal{AB}$ are defined as
\begin{align}
\mathcal{I}(\hat{\rho}_{\mathcal{AB}}) &= S(\hat{\rho}_{A}) + S(\hat{\rho}_{B}) - S(\hat{\rho}_{\mathcal{AB}}), \\
\mathcal{J}_{A}(\hat{\rho}_{\mathcal{AB}}) &= S(\hat{\rho}_{A}) - S(\hat{\rho}_{\mathcal{A}}|\hat{\rho}_{\mathcal{B}}),
\end{align}
\noindent where $S(\hat{\rho}) = -\text{Tr}(\hat{\rho} \log_2 \hat{\rho})$ is the von Neumann entropy of the state $\hat{\rho}$, $S(\hat{\rho}_{A})$ and $S(\hat{\rho}_{B})$ are, respectively, the entropies of the reduced density matrices of the subsystems $A$ and $B$, and $S(\hat{\rho}_{\mathcal{A}}|\hat{\rho}_{\mathcal{B}})$ is the conditional entropy of $\mathcal{A}$ given $\mathcal{B}$. Quantum discord is then defined as the difference between these mutual informations, $\mathcal{D}(\hat{\rho}_{\mathcal{AB}}) = \mathcal{I}(\hat{\rho}_{\mathcal{AB}}) - \mathcal{J}_{A}(\hat{\rho}_{\mathcal{AB}})$.  Since $\mathcal{J}_{A}(\hat{\rho}_{\mathcal{AB}})$ may take into account classical correlations, it must be maximized considering every possible set of projective measurements onto the eigenstates, in order to ensure the quantum discord computes only nonclassical correlations. This leads to
\begin{equation}
\mathcal{D}(\hat{\rho}_{\mathcal{AB}}) = S(\hat{\rho}_{A}) - S(\hat{\rho}_{\mathcal{AB}}) + \min_{\{\Pi_{j}^{\mathcal{A}}\}} S(\hat{\rho}_{\mathcal{B}|\{\Pi_{j}^{\mathcal{A}}\}}),
\end{equation}
\noindent where $S(\hat{\rho}_{\mathcal{B}|\{\Pi_{j}^{\mathcal{A}}\}})$ is the entropy conditioned by measurements on subsystem $A$. Quantum discord can be non-zero even for separable (non-entangled) states, distinguishing it from entanglement~\cite{luo2008local, Coto_2017, Ali_2010}. From now on, for the purposes of this work, when we cite mutual information, we will be referring to the definition of $\mathcal{I}(\hat{\rho}_{\mathcal{AB}})$.


\subsection{Quantum Coherence}
Quantum coherence is a fundamental property of quantum systems and plays a crucial role in quantum information processing. It is typically characterized by the off-diagonal elements of the density matrix in a chosen basis. One prominent measure of quantum coherence is the coherence relative entropy, denoted here by $C_{\text{RE}}$, which quantifies the distance between a state $\hat{\rho}$ and its nearest diagonal (``classical") state $\hat{\rho}_{\text{diag}}$. This is defined as
\begin{equation}
C_{\text{RE}} = S(\hat{\rho}_{\text{diag}}) - S(\hat{\rho}).
\end{equation}
\noindent The coherence relative entropy provides a measure of the quantum coherence in a state, with $C_{\text{RE}} = 0$ indicating a diagonal (classical) state, and nonzero values signifying the presence of quantum coherence~\cite{baumgratz2014quantifying, streltsov2017colloquium}.

It is important to emphasize that quantum coherence is basis-dependent. The coherence of a state can vary depending on the choice of measurement basis, meaning that different bases can reveal different aspects of the coherence. This basis-dependence is a key distinction between quantum coherence and classical correlations, the latter being invariant under changes in the measurement basis. Consequently, the measure of coherence on any given basis is not absolute but depends on the specific representation chosen for the system. In this study, we measure quantum coherence for the qubits in the $S_z$ basis, where the off-diagonal elements of the density matrix are most indicative of coherence~\cite{baumgratz2014quantifying, streltsov2017colloquium}.

\subsection{Local Quantum Uncertainty}
Local Quantum Uncertainty is a measure of the minimal quantum uncertainty achievable through a local measurement on a bipartite quantum system. It is based on the skew information, which quantifies the deviation from commutativity between the quantum state and an observable. The skew information ${\cal K}(\hat{\rho}, \hat{K}_A)$ is defined as
\begin{equation}
{\cal K}(\hat{\rho},\hat{K}_A)=-\frac 12 \text{Tr}\{[\hat{\rho}^{\frac 12},\hat{K}_A]^2\}, \label{skewinfo}
\end{equation}
where $\hat{K}_A$ is the observable on subsystem $A$ and $\hat{\rho}$ is the state of the bipartite system. This quantity is nonnegative and vanishes if and only if the state and the observable commute, providing a natural measure of the quantum uncertainty due to noncommutativity~\cite{Girolami_PRL_13}.

The LQU is defined as the minimum skew information over all local observables in subsystem \( A \) of a bipartite system \( \hat{\rho}_{AB} \), and is given by
\begin{equation}
{\cal U}_A(\hat{\rho}_{AB}) \equiv \min_{\hat{K}_A} {\cal I}(\hat{\rho}, \hat{K}_A).\label{eq:ss1}
\end{equation}
For a system where both subsystems \( A \) and \( B \) are qubits, the minimization in Eq.~\eqref{eq:ss1} can be expressed in closed form, yielding a computable expression for \( {\cal U}_A(\hat{\rho}_{AB}) \). Specifically, for qubit-qubit systems, the LQU can be computed as
\begin{equation}
{\cal U}_A(\hat{\rho}_{AB}) = 1 - \lambda_{\text{max}}\left\{W_{AB}\right\}, \label{anale}
\end{equation}
\noindent where \( \lambda_{\text{max}} \) is the maximum eigenvalue, and \( W_{AB} \) is a symmetric \( 3 \times 3 \) matrix with elements
\[
(W_{AB})_{ij} = \text{Tr}\left\{ \hat{\rho}_{AB}^{1/2} (\hat{\sigma}_i^A \otimes \mathbb{I}_B) \hat{\rho}_{AB}^{1/2} (\hat{\sigma}_j^A \otimes \mathbb{I}_B) \right\},
\]
where \( \hat{\sigma}_i^A \) are the Pauli matrices (\( i = x, y, z \)) acting on subsystem \( A \), and \( \mathbb{I}_B \) is the identity matrix acting on subsystem \( B \).

For pure states \( |\psi_{AB}\rangle \), this formula reduces to the linear entropy of entanglement
\[
{\cal U}_A(\hat{\rho}_{AB}) = 2\left(1 - \text{Tr}(\hat{\rho}_A^2)\right),
\]
where \( \hat{\rho}_A \) is the reduced density matrix of subsystem \( A \). In the case of maximally entangled pure states, the LQU is normalized to unity, reflecting maximal quantum uncertainty in the system~\cite{Girolami_PRL_13}.

\section{Results and discussion}\label{Results and discussion}

\begin{figure}[h!]
\begin{center}
\includegraphics[width=0.49\textwidth]{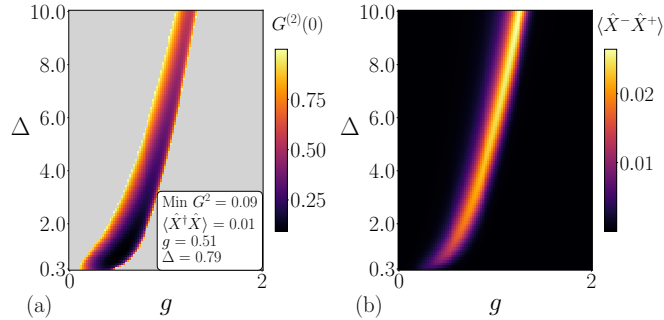}
\end{center}
\vspace{-0.7cm}\caption{
Heatmaps of (a) the second-order correlation function $G^{(2)}(0)$ and (b) the number of excitations $\langle \hat{X}^- \hat{X}^+ \rangle$ as functions of the coupling strength $g$ and the qubit frequency $\Delta$ for degenerate qubits ($\Delta_1 = \Delta_2 = \Delta$) and symmetric coupling ($g_1 = g_2 = g$) at a fixed temperature $\textbf{T} = 0.1$, with $\omega = 1$. Regions with $G^{(2)}(0) \geq 1$ are shown in light gray in panel (a). The minimum value for the second-order correlation function, $G^{(2)}(0) = 0.09$, is observed at $g = 0.51$ and $\Delta = 0.79$, corresponding to excitation number of $\langle \hat{X}^- \hat{X}^+ \rangle = 0.01$.}\label{g2}
\end{figure}

We begin by examining the quantum properties associated with photon behavior, focusing on the second-order correlation function $G^{(2)}(0)$, which is calculated using the dressed jump operator. Fig.~\ref{g2}(a) illustrates its dependence on degenerate coupling strengths ($g_1 = g_2 = g$) and transition frequencies ($\Delta_1 = \Delta_2 = \Delta$) for a fixed mode frequency $\omega = 1$ and temperature $\textbf{T} = 0.1$. Hereafter, we adopt natural units, expressing all parameters in terms of the lower frequency of the system. Specifically, we use the mode frequency in the high-frequency limit of the qubit and the qubit frequency in the regime of adiabatic quantum oscillators, as in Fig.~\ref{EnPn}.

Photon antibunching, characterized by $G^{(2)}(0) < 1$, emerges under specific spectral conditions in quantum systems governed by a general Hamiltonian. This phenomenon has been widely used to characterize the nonclassicality of emitted photons~\cite{ashhab2010qubit,shen2014ground,ridolfo2013nonclassical,garziano2017cavity}. Under thermal equilibrium, a critical requirement for achieving antibunching is the presence of a quasi-degenerate ground-state structure. In such cases, the ground state and the first excited state form a closely spaced doublet that is isolated from higher energy levels. This configuration ensures that, at low temperatures, thermal excitations remain confined primarily to these two states, minimizing the population in higher energy levels.

Symmetry plays a pivotal role in constraining transitions within the system. For the eigenstates of the Hamiltonian of Eq.~\eqref{eq:Ham}, the matrix elements $X_{jk} = \langle \phi_j | (\hat{a}^\dagger + \hat{a}) | \phi_k \rangle$ are restricted to transitions between states of differing parity. These symmetry constraints suppress multi-photon processes, thereby favoring single-photon emissions. Furthermore, the energy gap between the ground and first excited states ($\delta_{01}$) must be significantly smaller than the gaps to higher levels ($\delta_{12}, \delta_{23}, \ldots$). This energy hierarchy ensures that higher states remain thermally inaccessible or only weakly populated under typical low-temperature conditions, further reducing the likelihood of cascaded or multi-photon emissions. Nonlinearity in the spectrum also disrupts the resonance conditions that facilitate simultaneous multi-photon emissions, thereby favoring single-photon transitions~\cite{rabl2011photon,ridolfo2012photon}.

\begin{figure}[t!]
\begin{center}
\includegraphics[width=0.48\textwidth]{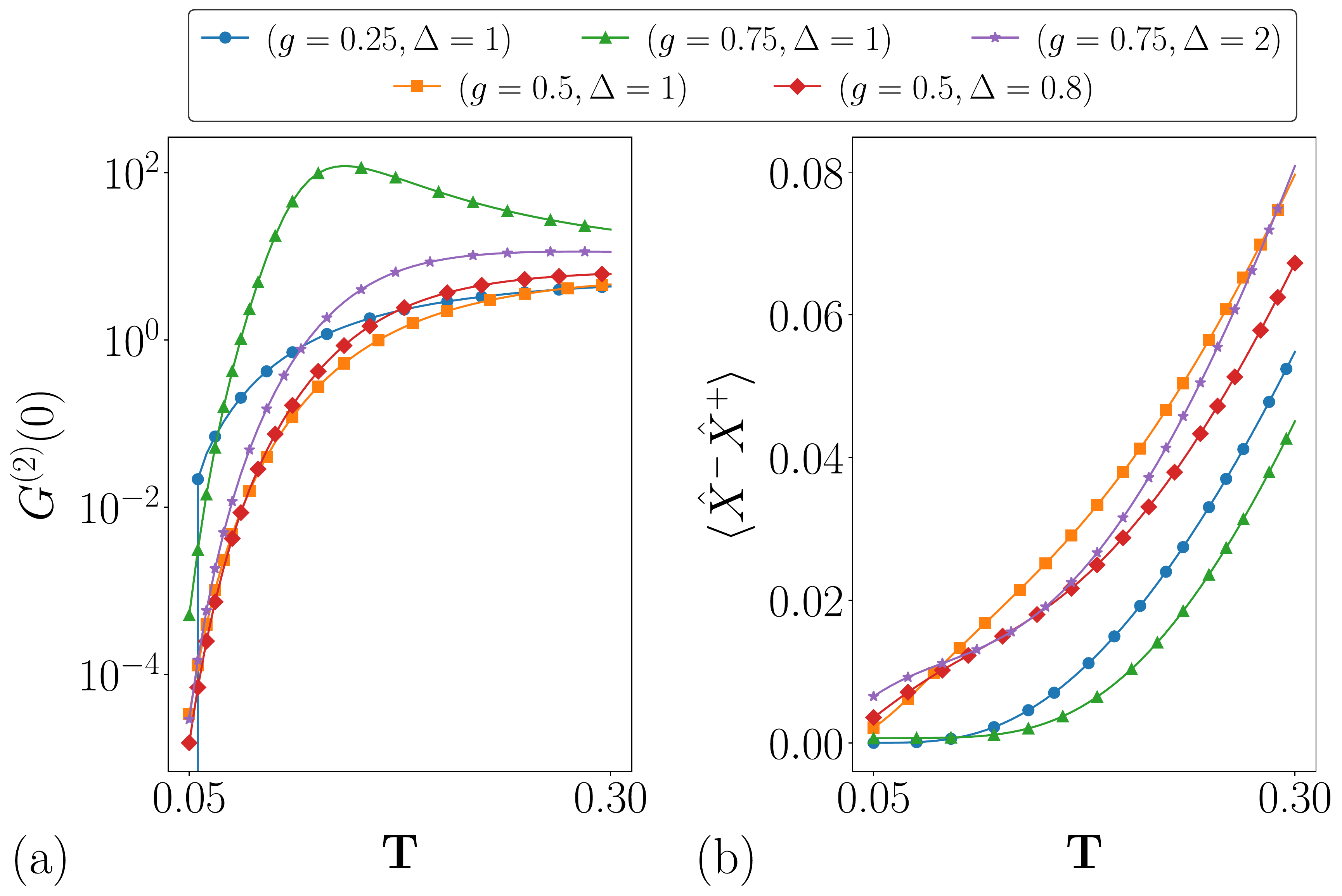}
\end{center}
\vspace{-0.7cm}\caption{(a) Temperature ($\textbf{T}$) dependence of the second-order correlation function $G^{(2)}(0)$ and (b) the number of excitation $\langle \hat{X}^- \hat{X}^+ \rangle$ for a fixed mode frequency $\omega = 1$, with degenerate qubit frequencies $\Delta \in [0.8, 1, 2]$ and symmetric coupling strengths $g \in [0.25, 0.50, 0.75]$. For $\textbf{T} < 0.1$, the second-order correlation function rapidly increases, while the number of excitations presents similar behavior for $\textbf{T} > 0.1$. }\label{G2T}
\end{figure}

The interplay of these factors -- quasi-degenerate states, symmetry constraints, and energy gap hierarchy -- is evident in the two-qubit quantum Rabi model, particularly near resonance in the ultrastrong coupling regime. This behavior is depicted in Figs.~\ref{EnPn}(b) and corroborated in Fig.~\ref{g2}(a).

In Ref.~\cite{nodar2023identifying}, $G^{(2)}(0)$ was used to identify the breakdown of the rotating wave approximation far below the conventional threshold of the ultrastrong coupling (USC) regime. The observed strong bunching behavior, characterized by $G^{(2)}(0) \gg 2$, was attributed to specific spectral features of the system. As shown in Fig.~\ref{g2}(a), regions where $G^{(2)}(0) > 1$ are highlighted in light gray, indicating that $G^{(2)}(0)$ cannot reliably serve as a nonclassicality quantifier in such cases, even when strong bunching occurs. Additionally, we note that antibunching-to-bunching transitions associated with first-order quantum phase transitions in the AQRM~\cite{ye2023implication,xu2024persisting} are absent here, consistent with observations in the standard QRM.
It is worth noting that earlier studies on the thermal QRM and its variants predominantly focused on the resonant case. However, as demonstrated in Fig.~\ref{g2}, detuning offers two significant advantages. First, it facilitates the attainment of lower values of $G^{(2)}(0)$ at relatively small coupling strengths $g$, as seen in Fig.~\ref{g2}(a). Second, it enhances the photon emission rate from the resonator, which is proportional to $\langle \hat{X}^- \hat{X}^+ \rangle$ and detectable by a photon-absorber~\cite{garziano2013switching}, as shown in Fig.~\ref{g2}(b). For the temperature under consideration, $\textbf{T} = 0.1$, the minimum value of the equal-time second-order correlation function is $G^{(2)}(0) = 0.09$, occurring at a coupling strength of $g = 0.51$ and a qubit frequency of $\Delta = 0.79$, with a photon number $\langle \hat{X}^- \hat{X}^+ \rangle = 0.01$.

Beyond two-photon correlations, higher-order photon correlation functions have also been extensively studied to provide a more accurate characterization of photon statistics. These higher-order functions are closely related to phenomena such as multi-photon blockade and $n$-photon sources~\cite{zubizarreta2020conventional,hamsen2017two}. However, for the given parameters at a temperature of $\textbf{T} = 0.1$, the condition for two-photon blockade, i.e., $G^{(3)}(0) < 1$ and $G^{(2)}(0) > 1$, is not satisfied. As the information provided by the equal-time third-order correlation function, $G^{(3)}(0)$, is identical to that of $G^{(2)}(0)$ regarding the nonclassicality of the system, it has been omitted from the analysis.

The computation of $G^{(2)}(0)$ using the bare bosonic field, when the coupling strength is turned off after the system reaches thermal equilibrium, results in strong photon bunching up to the USC regime. As the light-matter coupling transitions into the DSC regime, $G^{(2)}(0)$ approaches unity, regardless of the detuning. This indicates statistical properties characteristic of a coherent state present in the thermal emission. In this scenario, the 2QQRM behaves identically to the QRM, and thus, the corresponding plots are omitted.

\begin{figure}[b!]
\begin{center}
\includegraphics[width=0.49\textwidth]{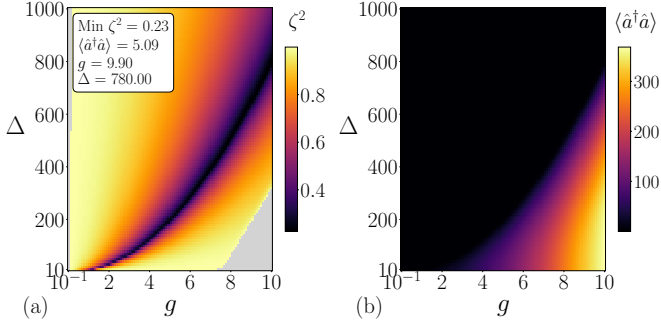}
\end{center}
\vspace{-0.7cm}\caption{
Heatmaps of (a) the squeezing parameter $\zeta^2$ and (b) the photon population $\langle \hat{a}^\dagger \hat{a} \rangle$ as functions of the symmetric coupling strength $g$ and the degenerate qubit frequencies $\Delta$, at a fixed temperature of $\textbf{T} = 0.1$ and field frequency $\omega = 1$. In panel (a), regions where $\zeta^2 \geq 1$ are shown in light gray. The minimum squeezing parameter, $\zeta^2 = 0.23$, occurs at $g = 9.9$ and $\Delta = 780$, corresponding to a photon population of $\langle \hat{a}^\dagger \hat{a} \rangle = 5.09$. Strong squeezing is observed just below the critical coupling line $g_c = \sqrt{\omega \Delta}/2$, beyond which the system enters the superradiant phase, characterized by an increased photon population and a decrease in squeezing.}\label{squeezing1}
\end{figure}

Fig.~\ref{G2T} shows the impact of temperature on the two-photon correlation function $G^{(2)}(0)$, presented in Fig.~\ref{G2T}(a), and the number of excitations $\langle \hat{X}^- \hat{X}^+ \rangle$, depicted in Fig.~\ref{G2T}(b), for various (symmetric) coupling strengths and (degenerate) qubit frequencies $\{g, \Delta\} = \{0.25, 1\}, \{0.5, 1\}, \{0.75, 1\}, \{0.5, 0.8\}, \{0.75, 2\}$. Notably, photon antibunching occurs only when $\textbf{T} < 0.2$, and identical $G^{(2)}(0)$ values can be achieved for varying excitation numbers by carefully tuning the parameters $\{g, \Delta\}$.

\begin{figure}[t!]
\begin{center}
\includegraphics[width=0.485\textwidth]{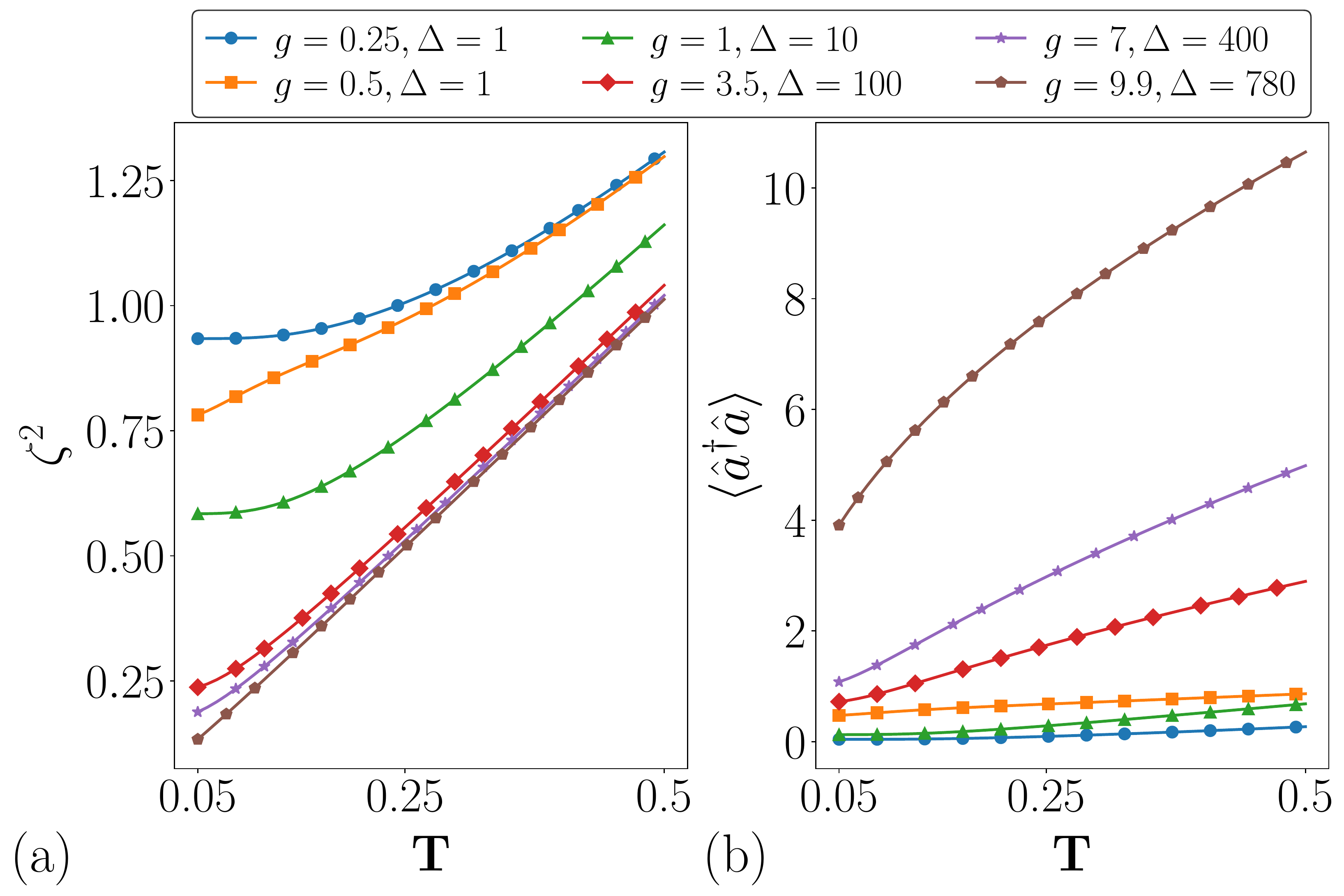}
\end{center}
\vspace{-0.7cm}\caption{Temperature ($\textbf{T}$) dependence of the squeezing parameter $\zeta^2$ (a) and photon population $\langle \hat{a}^\dagger \hat{a} \rangle$ (b) for a fixed mode frequency $\omega = 1$, with degenerate qubit frequencies $\Delta/\omega \in [1, 10, 100, 400, 780]$ and symmetric coupling strengths $g/\omega \in [1, 2, 3.5, 7, 9.9]$. For $\textbf{T} < 0.1$, the squeezing parameter remains nearly constant while the photon population increases. As temperature rises, squeezing decreases linearly and vanishes around $\textbf{T} \sim 0.5$, while the photon population continues to increase, especially at higher couplings. }\label{squeezingT}
\end{figure}

\begin{figure*}[t!]
\begin{center}
\includegraphics[width=1\textwidth]{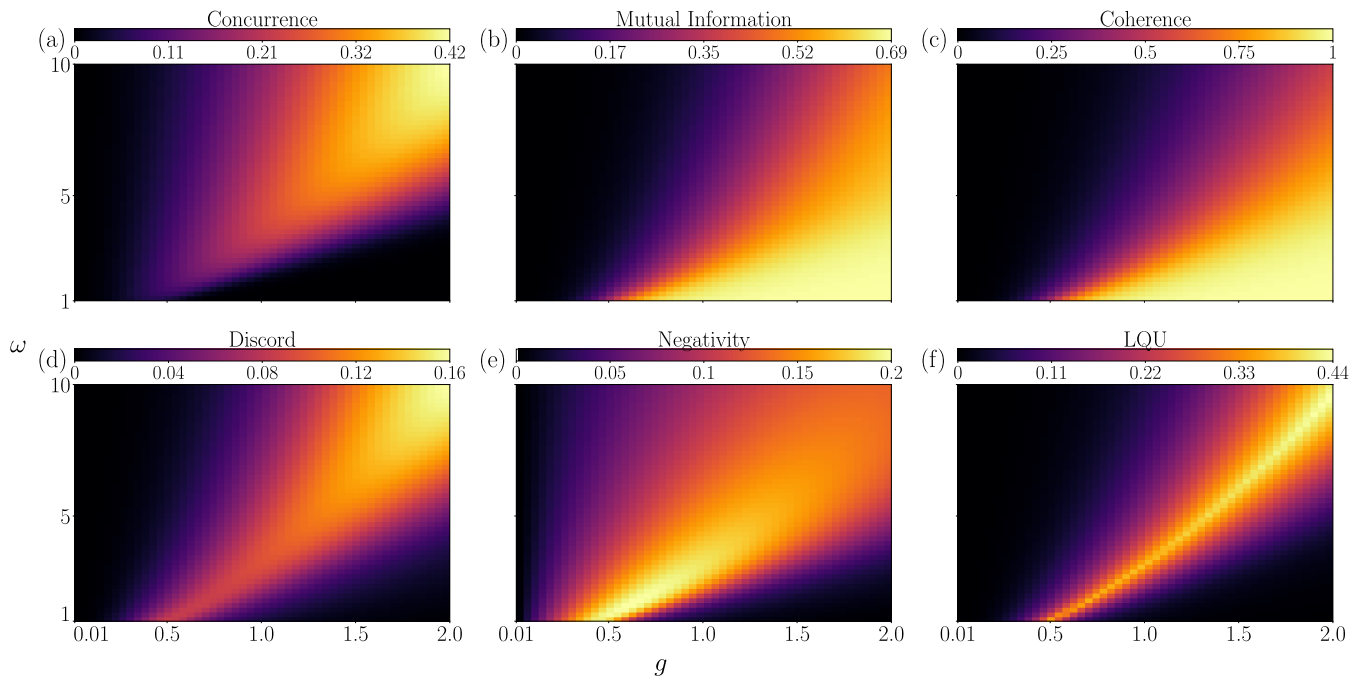}
\end{center}
\vspace{-0.7cm}\caption{Heatmaps showing nonclassicality quantifiers for the two qubits around resonance, as a function of the symmetric coupling strength $g$ and the mode frequency $\omega$, with a fixed temperature $\textbf{T}=0.1$. From top to bottom and left to right, the panels display (a) concurrence, (b) mutual information, (c) coherence, (d) quantum discord, (e) negativity, and (f) local quantum uncertainty (LQU). We set degenerate qubit frequencies $\Delta=1$ in all plots.}\label{NQ1}
\end{figure*}

To examine the degree of photonic squeezing, we present heatmaps of the squeezing parameter $\zeta^2$ in Fig.~\ref{squeezing1}(a) and the photon population $\langle \hat{a}^\dagger \hat{a} \rangle$ in Fig.~\ref{squeezing1}(b) as functions of symmetric coupling strengths $g_1=g_2=g$ and degenerate qubit frequencies $\Delta_1 = \Delta_2 = \Delta$, at a fixed temperature $\textbf{T} = 0.1$ and mode frequency $\omega=1$. In Fig.~\ref{squeezing1}(a), regions with $\zeta^2 \geq 1$ are shown in light gray, while areas with $\zeta^2 < 1$ indicate significant photonic squeezing, occurring for specific coupling strengths within the range of the qubit frequencies. From the effective Hamiltonian of Eq.~\eqref{eq:Heff}, squeezing is expected in the non-resonant regime. The minimum squeezing parameter, $\zeta^2 = 0.23$, is achieved at $g = 9.9$ and $\Delta = 780$, with a corresponding photon population of $\langle \hat{a}^\dagger \hat{a} \rangle = 5.09$. Additionally, pronounced squeezing occurs just below the critical coupling strength $g_c = \frac{\sqrt{\omega \Delta}}{2}$, although the system does not strictly adhere to the thermodynamic limit $\omega/\Delta \rightarrow 0$. At these points, the system remains in the normal phase, which supports quadrature squeezing. In contrast, beyond $g_c$, the system transitions into the superradiant phase, characterized by high photon populations and the absence of squeezing, as illustrated in Fig.~\ref{squeezing1}(b). For small $g$, squeezing increases with the coupling strength but remains weak near the qubit-photon resonance region. The strongest squeezing is observed in the high-frequency qubit regime, aligning with QRM ground-state results~\cite{ashhab2010qubit}. This behavior stems from the flattening of the oscillator’s effective potential near the critical point, which leads to a significant enhancement in squeezing.

In Fig.~\ref{squeezingT}, we explore the impact of temperature on both the squeezing parameter $\zeta^2$ and the photon population $\langle \hat{a}^\dagger \hat{a} \rangle$. Fig.~\ref{squeezingT}(a) shows the squeezing parameter $\zeta^2$, and Fig.~\ref{squeezingT}(b) displays the corresponding photon population as a function of temperature $\textbf{T}$, expressed in units of the mode frequency $\omega$. The analysis is performed for a fixed mode frequency $\omega = 1$, and a variety of qubit frequencies $\Delta_1 = \Delta_2 = \Delta \in [1, 10, 100, 400, 780]$, along with different coupling strengths $g_1 = g_2 = g \in [1, 2, 3.5, 7, 9.9]$. As observed in both panels of the figure, for low temperatures ($\textbf{T} < 0.1$), the squeezing parameter remains nearly constant, as does the number of photons, indicating that the system continues to exhibit squeezing despite the presence of temperature. However, as the temperature increases, a noticeable change occurs: the squeezing decreases approximately linearly with temperature for all sets of qubit frequencies and coupling strengths. This indicates how the thermal effects progressively reduce the degree of squeezing in the system. At temperatures around $\textbf{T} \sim 0.5$, the system loses its ability to exhibit squeezing. Additionally, the photon population increases with temperature, especially at higher coupling strengths, reflecting the enhanced thermal excitation of the system. These results highlight the delicate balance between squeezing and thermal excitation, where squeezing is more robust at low temperatures and diminishes as thermal effects become more pronounced. Interestingly, squeezing survives to moderate temperatures, though it is reduced as the system is subjected to increased thermal energy.
Furthermore, different couplings $g_1$ and $g_2$ and qubit frequencies $\Delta_1$ and $\Delta_2$ were studied, but no significant improvement was found in the achievable degree of squeezing, so these variations are omitted from the discussion.

We now examine the quantum correlations between the qubit-qubit, as well as between the field and the qubit. Fig.~\ref{NQ1} presents heatmaps of these correlations around resonance as functions of the coupling strength $g$ (ranging from 0 to 2) and the field transition frequency $\omega$ (ranging from 0.1 to 10) at a fixed temperature of $\textbf{T} = 0.1$ and degenerate qubit frequencies $\Delta=1$. Among the quantum correlation measures, negativity (Fig.~\ref{NQ1}(e)) is the weakest, remaining below $\mathcal{N}(\hat{\rho}_\mathrm{qubits}) = 0.2$, even under optimal conditions where qubit and field frequencies are close in the USC regime. In contrast, other measures exhibit significant nonclassical signatures as the coupling strength increases and the system transitions into the DSC regime. In this case, mutual information (Fig.~\ref{NQ1}(b)) and coherence (Fig.~\ref{NQ1}(c)) attain their highest values near resonance. Conversely, the best results for concurrence (Fig.~\ref{NQ1}(a)) and quantum discord (Fig.~\ref{NQ1}(d)) are found for large detunings. A similar behavior is observed for local quantum uncertainty (LQU, Fig.~\ref{NQ1}(f)), but it remains confined to a narrow range.

Notably, coherence reaches its theoretical upper bound $(C_\mathrm{RE} = 1)$, as expected from Eq.~\eqref{e.unit_trans} in the adiabatic approximation. In this regime, the system’s eigenstates are product states of displaced Fock states for the photons and $x$-polarized atomic states, $\ket{\Phi} = \ket{j,m} \ket{\phi_{m}}$, leading to vanishing qubit-photon entanglement. Mutual information is also substantial, approaching $\mathcal{I}(\hat{\rho}_\mathrm{qubits}) \approx 0.7$, while concurrence and LQU reach moderate values, around $\mathcal{C}(\hat{\rho}_\mathrm{qubits}) \sim \mathcal{U}(\hat{\rho}_{\mathrm{qubits}}) \sim 0.4$. These results align with Ref.~\cite{abari2024correlated}, which analyzed two ideal qubits interacting with a thermal-mechanical mode under resonance. However, our study extends this framework by exploring the impact of DSC at large detunings within a fully thermalized equilibrium state, a regime that has not been previously investigated.

\begin{figure*}[t!]
\begin{center}
\includegraphics[width=1\textwidth]{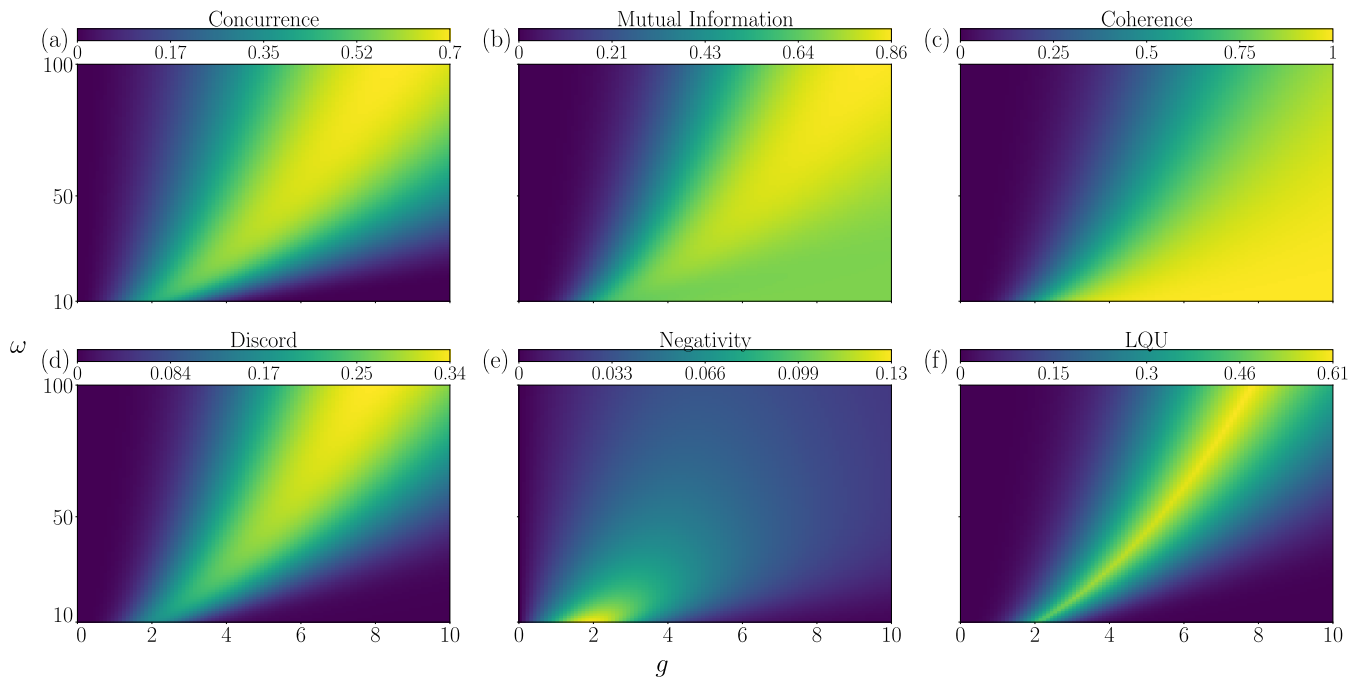}
\end{center}
\vspace{-0.7cm}\caption{Heatmaps displaying nonclassicality quantifiers for the two-qubit non-resonant scenario as functions of the symmetric coupling strength $g$ and the field transition frequency $\omega$, with a fixed temperature of $\textbf{T} = 0.1$. The panels, arranged from top to bottom and left to right, show: (a) concurrence, (b) mutual information, (c) coherence, (d) quantum discord, (e) negativity, and (f) local quantum uncertainty (LQU). All plots assume degenerate qubit frequencies with $\Delta = 1$. }\label{NQ2}
\end{figure*}

Fig.~\ref{NQ2} deepens our analysis by systematically exploring the effects of detuning over a broader parameter range. In non-resonant regimes, Eqs.~\eqref{H disp} and~\eqref{eq:Heff} reveal the emergence of effective qubit-qubit interactions mediated by their dispersive coupling to the field. Specifically, Eq.~\eqref{eq:Heff} describes the non-RWA dispersive regime ($g \ll |\Delta - \omega|$), where strong quantum correlations are more favorably established. While coherence, Fig.~\ref{NQ2}(c), follows the same pattern observed previously, other correlation measures exhibit distinct changes. Most notably and also in agreement with the predictions from Eqs.~\eqref{H disp} and~\eqref{eq:Heff}, negativity (Fig.~\ref{NQ2}(e)) experiences a significant reduction -- approximately halved -- despite increasing the coupling strength up to $g=10$. Conversely, larger detunings enhance the remaining quantum correlation quantifiers, with all optimal values occurring at higher field frequencies. Quantum discord, Fig.~\ref{NQ2}(d), exhibits the most pronounced improvement, more than doubling compared to Fig.~\ref{NQ1}(d), although it remains weaker than the other correlations, only surpassing negativity. Meanwhile, concurrence (Fig.~\ref{NQ2}(a)) and LQU (Fig.~\ref{NQ2}(f)) reach remarkable values, with $\mathcal{C}(\hat{\rho}_\mathrm{qubits}) \sim 0.7$ and $\mathcal{U}(\hat{\rho}_{\mathrm{qubits}}) \sim 0.6$, respectively. Mutual information, Fig.~\ref{NQ2}(b), also increases by approximately 10\%, attaining $\mathcal{I}(\hat{\rho}_\mathrm{qubits}) \approx 0.86$. These trends underscore the role of detuning in amplifying quantum correlations, particularly in the dispersive DSC regime.

\begin{figure*}
\begin{center}
\includegraphics[width=1\textwidth]{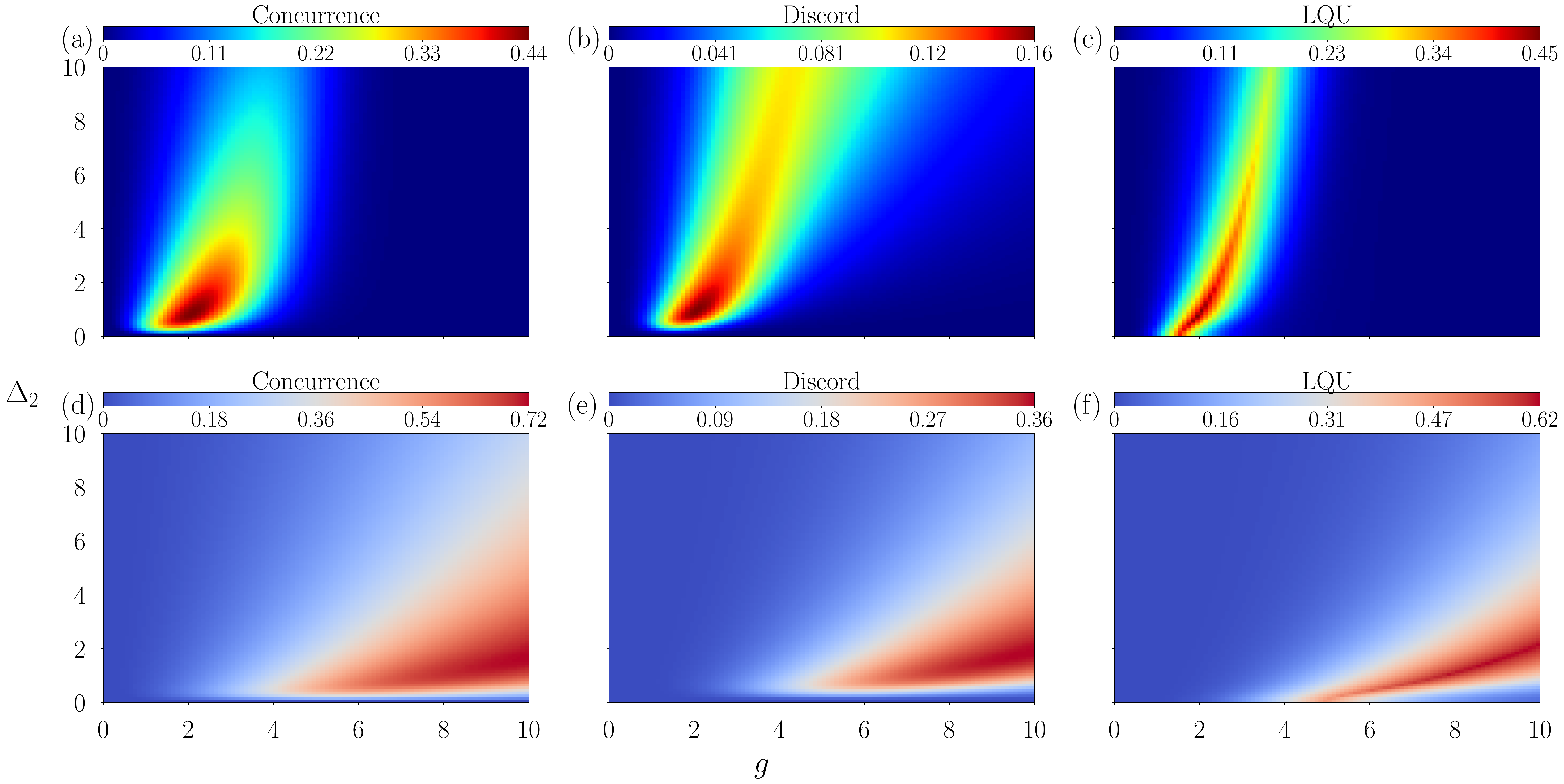}
\end{center}
\vspace{-0.7cm}\caption{Heatmaps showing the concurrence (a)-(d), quantum discord (b)-(e), and local quantum uncertainty (LQU) (c)-(f) as functions of the coupling strength $g$ and the transition frequency $\Delta_2$ of the second qubits. The field frequency is set to $\omega = 10$ in the top panels and $\omega = 100$ in the bottom panels. The temperature is held constant at $\textbf{T} = 0.1$. In all cases, we set the frequency of the first qubit as $\Delta = 1$. }\label{NQ_delta2}
\end{figure*}

To further enhance the nonclassical features of the two qubits and assess their robustness against parameter fluctuations, Fig.~\ref{NQ_delta2} qualitatively explores the effects of varying qubits' detunings on the most sensitive quantities, namely, concurrence (Figs.~\ref{NQ_delta2}(a)-(d)), quantum discord (Figs.~\ref{NQ_delta2}(b)-(e)), and local quantum uncertainty (Figs.~\ref{NQ_delta2}(c)-(f)). Specifically, we fix the transition frequency of the first qubit at $\Delta_1=1$ and analyze two cases for the mode frequency: (i) $\omega = 10$ (top panels) and (ii) $\omega = 100$ (bottom panels). At first, it is worth noticing that the coupling strength $g$ plays a decisive role in the emergence of nonclassical effects, as these effects are suppressed in certain parameter regimes, specifically, at higher couplings in the top panels and at lower couplings in the bottom panels. In the top panels, where the field frequency is closer to the transition frequency of the first qubit, quantum correlations are enhanced at smaller values of $g$, allowing them to emerge even for varying $\Delta_2$. Conversely, in the bottom panels, when $\omega \gg \Delta_1$ (bottom panels), nonclassical effects strengthen with large coupling strength, $g>4$.

Although the specific values of $g$ influence the prominence of quantum correlations and their dependence on the second qubit’s transition frequency, the maximum values occur for near-detuned qubits in both scenarios. Moreover, for a wide range of couplings, near-optimal results are achieved, highlighting the robustness of the system against parameter variations while preserving its quantum features.

\begin{figure}
\begin{center}
\includegraphics[width=0.495\textwidth]{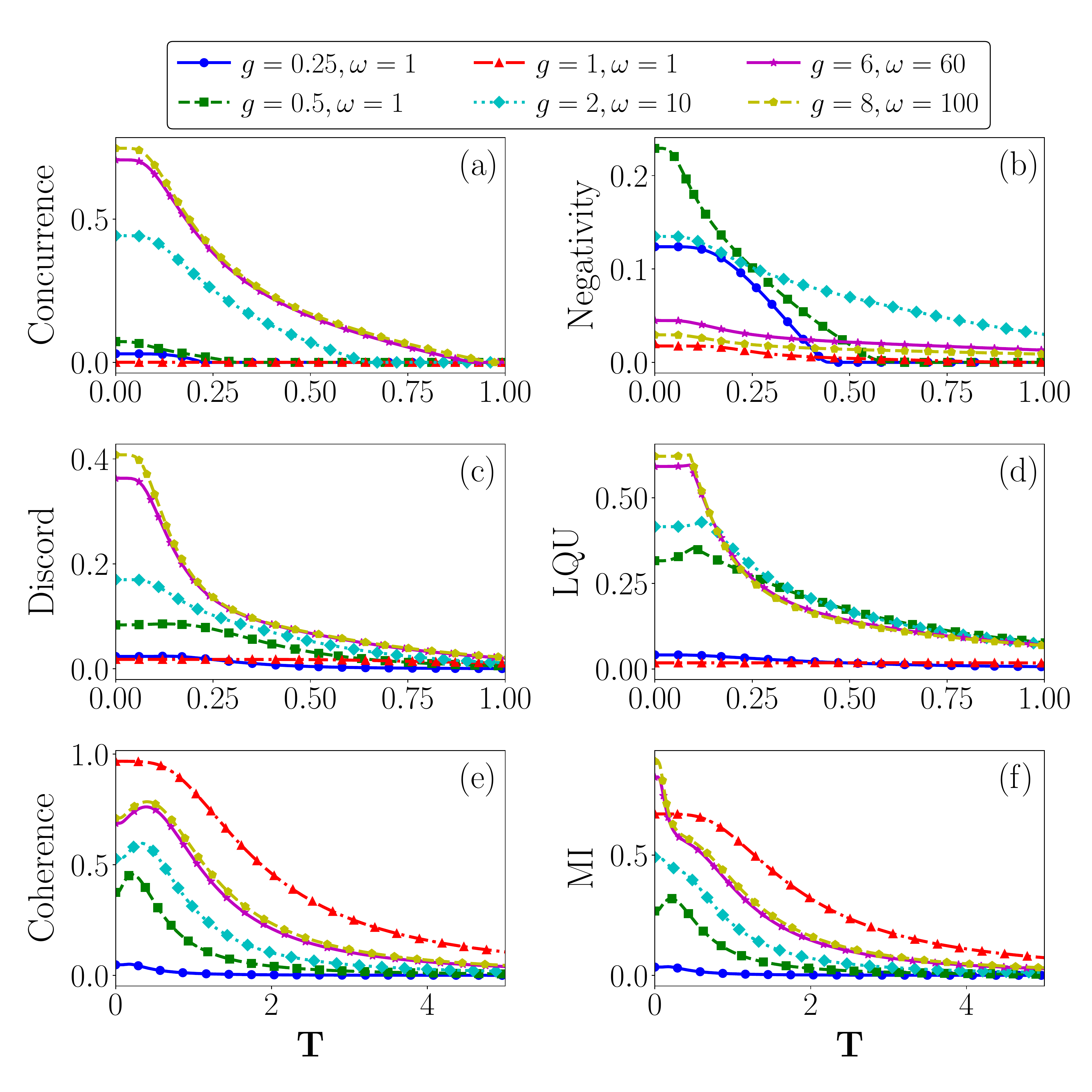}
\end{center}
\vspace{-0.7cm}\caption{Temperature ($\textbf{T}$) dependence of (a) concurrence, (b) negativity, (c) quantum discord, (d) local quantum uncertainty (LQU), (e) coherence, and (f) mutual information (MI) for a fixed qubit frequency $\Delta_1 = \Delta_2 = \Delta = 1$, with mode frequencies $\omega \in [1, 10, 60, 100]$ and coupling strengths $g \in [0.25, 0.5, 1.0, 2.0, 6.0, 8.0]$. For temperatures $T < 0.1$, all quantities remain nearly constant. As the temperature increases, for $T > 1.0$, only mutual information and coherence exhibit significant values. }\label{NQT}
\end{figure}

In Fig.~\ref{NQT}, we examine the impact of temperature on the nonclassical properties of the two qubits. From top to bottom and left to right, we present (a) concurrence, (b) mutual information (MI), (c) coherence, (d) quantum discord, (e) negativity, and (f) local quantum uncertainty (LQU) as functions of temperature $\textbf{T}$. The analysis is performed for a fixed (degenerate) transition frequency $\Delta = 1$ of the qubits, considering a range of field frequencies $\omega \in [1, 10, 60, 100]$ and coupling strengths $g \in [0.25, 0.5, 1.0, 2.0, 6.0, 8.0]$.

Consistent with the observations in Fig.~\ref{squeezingT}, the quantum properties of the qubits remain nearly constant at low temperatures ($\textbf{T} \lesssim 0.1$), regardless of the chosen parameters. However, at higher temperatures, the behavior of different quantities diverges. Specifically, concurrence (Fig.~\ref{NQT}(a)) and quantum discord (Fig.~\ref{NQT}(c)) decay to zero, following a trend similar to the squeezing parameter and second-order correlation function. While negativity (Fig.~\ref{NQT}(b)) and local quantum uncertainty (LQU, Fig.~\ref{NQT}(d)) also decrease significantly, they remain measurable for certain parameter sets, even at $\textbf{T} = 1$. In contrast, mutual information (Fig.~\ref{NQT}(f)) generally vanishes rapidly with increasing temperature across most parameter configurations, with the exception of the resonant case where $g = \omega = 1$, where it survives up to $\textbf{T} = 5$. Notably, at lower temperatures, some parameter sets (e.g., $g = \omega = 1$) yield nonzero mutual information while simultaneously exhibiting zero quantum discord, indicating that the correlations are predominantly classical in this regime. Conversely, parameter sets that produce both significant mutual information and quantum discord (e.g., $g = 6$ and $\omega = 60$) confirm the persistence of quantum correlations within the system.

Lastly, coherence (Fig.\ref{NQT}(e)) displays remarkable resilience to thermal effects. Interestingly, it initially increases with temperature before gradually diminishing at higher values of $\textbf{T}$. This counterintuitive behavior aligns qualitatively with recent findings\cite{abari2024correlated}, although their analysis considered thermal equilibrium only for the field mode, not the qubits. The observed enhancement reveals a regime where moderate thermal excitations can amplify coherence—a trend similarly seen in mutual information (MI) and local quantum uncertainty (LQU). While conventional nonclassicality measures such as entanglement and squeezing typically decay rapidly and exhibit strong quantum features only at zero temperature, these correlation-based quantifiers persist strikingly well. Beyond the optimal temperature point, coherence, MI, and LQU degrade much more slowly than other quantum resources across a wide range of parameters. This distinctive "thermal persistence" effect could offer significant advantages for applications demanding robustness against thermal noise.

\begin{table*}[t]
  \centering
  \caption{Comparative analysis of nonclassicality metrics between the two-qubit quantum Rabi model (2QQRM) and the standard QRM.}
  \label{tab:comparison}
  \begin{tabular}{|p{3.2cm}|p{13.5cm}|}
    \hline
    \textbf{Metric} & \textbf{New Insights from the 2QQRM} \\
    \hline
    $G^{(2)}(0)$ & Reveals stronger antibunching at lower coupling strengths compared to QRM. The most nonclassical photon statistics emerge at approximately half the coupling strength of the QRM. At higher coupling, the statistics approach those of coherent states. Detuning serves as a powerful control parameter -- its role was underexplored in prior QRM studies. \\
    \hline
    Squeezing & Observed beyond the resonant case, especially near the critical point in the USC regime. Unlike earlier QRM analyses limited to resonance, the 2QQRM unveils previously unreported squeezing behavior in off-resonant and critical regimes. \\
    \hline
    Negativity & Pronounced near resonance in the USC regime, but decreases when moving into the dispersive or DSC regime. Reflects strong but shifting qubit-field entanglement as coupling increases. The behavior is qualitatively similar to QRM. \\
    \hline
    Concurrence & Qubit-qubit entanglement is absent in the single-qubit QRM but emerges naturally in 2QQRM due to indirect interaction via the shared field. Especially prominent in the DSC regime under the high-frequency qubit condition and robust at moderate temperatures. \\
    \hline
    Quantum Discord & Tracks with concurrence, showing robust nonclassical correlations across broad parameter ranges. Useful when entanglement vanishes but quantum correlations persist. \\
    \hline
    Local Quantum Uncertainty (LQU) & Exhibits high values near resonance and in the USC regime, with detuning amplifying its magnitude in the transition to DSC. While prior QRM studies focused on nonequilibrium or closed scenarios, this work explores its thermal equilibrium behavior. \\
    \hline
    Quantum Coherence & Coherence unexpectedly increases at low temperatures, contrasting with typical decoherence behavior in QRM. Highlights the interplay between thermal effects and effective qubit-qubit interactions in the 2QQRM. \\
    \hline
  \end{tabular}
\end{table*}

\begin{table}[t]
\centering
\caption{Experimental platforms in the USC/DSC regime. }
\label{tab:experiments}
\renewcommand{\arraystretch}{1.3}
\scriptsize
\begin{tabular}{@{}lccccc@{}}
\toprule
\textbf{Platform (Ref.)} & $\omega/2\pi$ & $g/\omega$ & $T$ & $k_BT/\hbar\omega$ & \textbf{Switching} \\
\midrule
Supercond.~\cite{Yoshihara2017} & 4.5 GHz & 1.34 & 10 mK & 0.046 & 1--100 ns \\
Semiconductor~\cite{forn2019ultrastrong} & 363 THz & 143 & 10 mK & $9.1\!\times\!10^{-7}$ & 1--100 ps \\
Trapped ions~\cite{Koch2023} & 1 MHz & 650 & 1 mK & 3.3 & 1--10 $\mu$s \\
Ultracold atoms~\cite{Dareau2018} & 100 Hz & 100 & 1 $\mu$K & 3.3 & 1--100 $\mu$s \\
Cavity QED~\cite{Kockum2019} & 5 GHz & 0.5 & 15 mK & 0.066 & 0.1--10 ns \\
\bottomrule
\end{tabular}
\end{table}

\section{CONCLUDING REMARKS}\label{Conclusions}
In this study, we investigate the steady state of the 2QQRM by solving a dressed master equation at thermal equilibrium, which remains valid for arbitrary light-matter coupling strengths and yields a Gibbs state. Our key findings are as follows: (i) the emergence of quantum correlations and nonclassicality at thermal equilibrium. We show that quantum effects persist over a wide range of detunings and coupling strengths, even at moderate thermal temperatures, as confirmed by multiple measures of quantum correlation and nonclassicality; (ii) these quantum features are most prominent in the USC regime, particularly near the onset of the DSC regime and around criticality -- where thermalization times may diverge -- while they practically vanish in the deeper DSC regime; (iii) detuning plays a crucial role in enhancing nonclassicality and regulating the number of excitations. Furthermore, our results exhibit robustness across a broad parameter space without requiring fine-tuning of temperatures, frequencies, or coupling strengths.

The 2QQRM retains key nonclassical features of the QRM -- particularly photon antibunching, though now peaking at half the characteristic coupling strength -- while exhibiting several novel quantum behaviors. Although squeezing has been observed in the QRM in the high-frequency regime away from criticality at zero temperature \citep{ashhab2010qubit} or under resonant conditions in thermal equilibrium \citep{xu2024persisting}, we demonstrate that the 2QQRM displays markedly different behavior: strong squeezing emerges near criticality but becomes completely suppressed in the superradiant phase. Most importantly, we uncover three distinctive phenomena in the 2QQRM that represent either qualitative departures from or entirely new features compared to the single-qubit case: (i) diverging thermalization timescales; (ii) persistent entanglement between qubits (a feature fundamentally impossible in the single-qubit QRM); and (iii) robust quantum coherence with finite local quantum uncertainty -- none of which have been systematically investigated for the QRM under thermal equilibrium conditions to date.
 For clarity, we compile in Table~\ref{tab:comparison} a comparative summary of the nonclassicality measures explored in this study. This table distills the main findings into an accessible format, highlighting both shared and distinctive features of the two models.

Our theoretical proposal to achieve nonclassical equilibrium states is both fundamentally significant and experimentally feasible, as recent advances in strong coupling and low-temperature techniques demonstrate. The USC regime, characterized by $g/\omega > 0.1$, has been experimentally realized for more than a decade~\cite{anappara2009signatures} and continues to be realized in several systems~\cite{forn2019ultrastrong}. The DSC regime, with $g/\omega \approx 1.34$ and a temperature of approximately 45 mK~\cite{yoshihara2017superconducting}, is now within reach, with thermal entanglement observed in the quantum Rabi model. Crucially, these implementations also permit active control of the light-matter coupling: superconducting circuits achieve ns-scale switching via magnetic flux tuning, while semiconductor systems enable ultrafast ps-scale control through optical pulses. 

Deep-strong coupling considering the QRM is also observed with a single-trapped ion and, more recently, with trapped atoms~\cite{cai2021observation}, remarkably achieving couplings of $\sim 6.5$ greater than the field frequency~\cite{Koch2023}. In these atomic systems, the coupling can be modulated on $\mu$s timescales through trap potential adjustments or laser control. Additionally, novel fluxonium-based superconducting devices are being developed to detect virtual photons in the USC regime~\cite{giannelli2024detecting}, with the added capability of in situ coupling strength modulation at nanosecond speeds. Hybrid semiconductor-superconductor platforms further enhance flexibility in the implementation of the QRM~\cite{van2018microwave}, combining fast switching with robust coherence properties.

The switching capabilities across platforms create new opportunities for dynamical control of quantum effects: from ultrafast (ps-scale) modulation in semiconductor systems ideal for studying non-equilibrium phenomena, to the more gradual ($\mu$s-scale) control in atomic systems suitable for observing thermalization processes. This underscores the exciting potential to realize nonclassical equilibrium states within these coupling regimes. Finally, USC is also reached using cold atoms~\cite{Dareau2018} and metamaterials~\cite{Bayer2017}, each offering distinct approaches to coupling control -- from optical manipulation in atomic systems to microwave tuning in metamaterials.

 To contextualize these theoretical results and demonstrate the feasibility of implementation, we provide in Table~\ref{tab:experiments} an overview of key experimental platforms that have reached the ultrastrong and deep-strong coupling regimes, including their characteristic switching timescales. The table includes details on relevant parameters such as coupling ratios, operating temperatures, and quantum system types, thereby serving as a practical reference for identifying systems where the predicted nonclassical equilibrium states of the 2QQRM may be observed. This comparison highlights the growing experimental capacity to probe strong light-matter interactions under thermal equilibrium, bridging theoretical proposals with cutting-edge technologies.

\begin{acknowledgments}
This work was supported by the Coordenação de Aperfeiçoamento de Pessoal de Nível Superior - Brasil (CAPES) - Finance Code 001, National Council for Scientific and Technological Development (CNPq), Grant 311612/2021-
0, and by the S\~ao Paulo Research Foundation (FAPESP), Grant 2022/00209-6 and Grant 2022/10218-2.

\end{acknowledgments}

\bibliographystyle{unsrt}
\bibliography{bibliography.bib}

\end{document}